\documentclass[letterpaper]{JHEP3}
\usepackage{epsfig}

\newcommand{\roughly}[1]{\mathrel{\raise.3ex\hbox{$#1$\kern-0.85em
\lower1ex\hbox{$\sim$}}}}

\newcommand{\lsim}{\roughly<}

\def\exd{{\hbox{d}}}
\def\hg{\hat{g}}
\def\cg{\check{g}}

\def\ba{\begin{eqnarray}}
\def\ea{\end{eqnarray}}
\def\be{\begin{equation}}
\def\ee{\end{equation}}

\def\ssB{{\scriptscriptstyle B}}
\def\ssE{{\scriptscriptstyle E}}
\def\ssM{{\scriptscriptstyle M}}
\def\ssN{{\scriptscriptstyle N}}
\def\ssP{{\scriptscriptstyle P}}
\def\ssQ{{\scriptscriptstyle Q}}
\def\ssR{{\scriptscriptstyle R}}
\def\ssT{{\scriptscriptstyle T}}

\def\ssZ{{\scriptscriptstyle Z}}

\def\L{\mathcal{L}}

\def\eff{{\rm eff}}
\def\reg{{\rm reg}}
\def\GH{{\scriptscriptstyle GH}}

\def\nn{\nonumber}

\def\d{\mathrm{d}}

\def\({\left(}
\def\){\right)}

\def\hg{\hat{g}}

\def\pref#1{(\ref{#1})}

\title{Technical Naturalness on a Codimension-2 Brane}

\author{C.P. Burgess,${}^{1-3}$ D. Hoover,${}^4$ and
G. Tasinato${}^5$ \\
${}^1$ Perimeter Institute for Theoretical Physics, Waterloo ON,
N2L 2Y5, Canada.\\
${}^2$
Physics \& Astronomy, McMaster University, Hamilton ON, L8S 4M1,
Canada. \\
${}^3$ Theory Division, CERN, CH-1211 Geneva 23, Switzerland. \\
${}^4$ Physics Dept., McGill University, Montr\'eal, QC, H3A 2T8,
Canada.\\
${}^5$ Institut f\"ur Teoretische Physik, Universit\"at
Heidelberg, D-69120 Heidelberg, Germany
}

\date{}

\abstract { We compute how threshold effects obtained by
integrating out a heavy particle localized on a codimension-2
brane influence the properties of the brane and the bulk fields it
sources in $D=d+2$ dimensions. We do so using a recently developed
formalism for matching the characteristics of higher codimension
branes to the properties of the bulk fields they source. We show
that although the dominant heavy-mass dependence induced in the
low-energy codimension-2 tension has the generic size expected,
$T_2 \propto M^d$, the very-low-energy effective potential
governing the on-brane curvature once bulk KK modes are integrated
out can be additionally suppressed, by factors of order $\kappa^2
M^d$, where $\kappa$ is the bulk gravitational coupling. In the
special case of a codimension-2 brane in a 6D supersymmetric bulk
we also estimate the size of the contributions of short-wavelength
bulk loops near the brane, and find these can be similarly
suppressed.}

\begin{document}

\section{Introduction}

Much of modern thinking in particle physics about what should be
expected to replace the Standard Model at LHC energies is driven
by the idea that the Standard Model is an effective description of
some unknown, more fundamental, theory describing physics at
shorter distance scales, $\lambda = 1/M$, than we can presently
measure. This picture captures much of what makes the Standard
Model most attractive: it consists of the most general set of
interactions that are possible among the observed particles (plus
the Higgs boson) that involve only couplings having (engineering)
dimension (mass)${}^d$ for $d \ge 0$ \cite{SM}. This is just what
one would expect to describe any physics in the more fundamental
theory that is unsuppressed by powers of $1/M$.

What does {\it not} fit easily into this picture, however, are the
only two interactions allowed by the model that have dimensionful
couplings:
\be \label{SMrelevant}
 \L_{\rm rel} = -\sqrt{-g} \, \left[ m^4
 - \mu^2 H^\dagger H \right] \,,
\ee
where $H$ is the Higgs doublet.\footnote{We take a broad-minded
point of view, and include the couplings of the metric in what we
call the Standard Model, as is also consistent with the modern
interpretation of General Relativity also as a low-energy
effective field theory.} The problem with these is that agreement
with observations requires the scales $m$ and $\mu$ to be much
smaller than $M$, unlike what usually happens in low-energy
effective theories. Since such suppression of so-called relevant
operators is unusual, this difficulty is made into a virtue by
using it as a clue to guide our search for whatever the new
physics is that ultimately replaces the Standard Model. Since both
of the terms in eq.~\pref{SMrelevant} arise in the scalar
potential, one is led by these kinds of considerations to regard
systems that can handle and suppress contributions to scalar
potentials as particularly interesting candidates for the Standard
Model's short-distance (UV) completion. All of the most promising
theories proposed so far --- supersymmetric theories, models
without scalar fields and extra-dimensional scenarios --- are of
this type.

It is the purpose of this paper to try to understand in more
detail one of the remarkable ways extra-dimensional models can
suppress ultra-violet contributions to scalar potentials. As as
been noticed by many authors --- first within the context of
cosmic string back-reaction \cite{cosmicstrings} in four
dimensions, and then again for brane-world models in codimensions
one \cite{5DSelfTune} and two \cite{CG,SLED} --- extra-dimensional
field equations allow codimension two branes in extra dimensions
to have precisely flat induced geometries. This, despite having
significant nonzero homogeneous energy densities (or tensions),
and being coupled to higher dimensional gravity. By contrast, if
there were no extra dimensions, a nonzero space-filling constant
energy density would inevitably curve the geometry of spacetime
when coupled to gravity. This observation that the induced brane
geometry can be decoupled from its on-brane energy density
provides one of the very few potential ways forward for
understanding how it is that the observed acceleration of the
universe points to an energy density, $m^4$, with $m$ so much
smaller than almost all of the other scales found in the Standard
Model \cite{XDCC}.

Although the existence of higher dimensional solutions whose 4D
curvature is decoupled from the brane tension is suggestive, what
has been missing to date is a quantitative study of precisely how
(or if) the scalar potential in the low-energy effective theory
manages to remain insensitive to the integrating out of
high-energy scales. In particular if a higher dimensional scalar
field couples to the brane in addition to gravity, it is important
to understand under what circumstances the low energy action
describing the system has interesting special properties, similar
to the ones previously mentioned in the case of pure gravity. In
this paper we provide part of this missing analysis, by explicitly
integrating out (at one loop) a very massive brane field on a
codimension-2 brane, to see how this affects the low-energy
effective theory. We use for these purposes a scalar tensor theory
in a $D = d+2$-dimensional bulk coupled in a fairly generic way to
a $d$-dimensional codimension-2 brane, for which the matching
rules between brane properties and near-brane bulk asymptotics
have recently been worked out, following an effective approach, in
\cite{cod2ren}.

We find the following results
\begin{itemize}
\item Integrating out a brane field of mass $M$ generically
contributes an amount $M^d$ to the tension,\footnote{The subscript
`2' here is meant to emphasize that it is the tension of a
codimension-2 brane, and not of the regularizing codimension-1
brane that is introduced at intermediate points in the
analysis.} $T_2$, of a space-filling codimension-2 brane in
$d+2$ dimensions, and so is not suppressed relative to naive
expectations.
\item This tension does not necessarily imply a similarly large
contribution to the effective potential, $U_2$, in the
$d$-dimensional effective theory that governs the spacetime
curvatures at energies well below the Kaluza-Klein scale. When the
bulk is integrated out at the classical level, these results are
consistent with the existence, in
 extra-dimensional theories,
of flat solutions with nonzero
tensions.
\item By examining theories with scalar fields in the bulk we are
able to see that the situations where the low-energy curvatures
can be small are also those for which the codimension-2 brane has
little or no coupling to the bulk scalar, in agreement with the known
situations where large tensions coexist with flat on-brane
geometries.
\item In order to contrast the behaviour of codimension-2 sources
with those of the better-studied codimension-1 branes, we use a
representation of the codimension-2 brane in terms of a small
regularizing codimension-1 brane that encircles the position of
the codimension-2 object at a small radius
$\rho_b$~\cite{UVCaps,cod2ren}. From the point of view of the
low-energy effective theory, on scales much larger than $\rho_b$,
the main difference between such a regularizing brane and a
macroscopic codimension-1 brane is that the radius $\rho_b$ is not
a macroscopically observable variable, and it will therefore be
integrated out. As we will explain, at the classical level this
amounts to self-consistently determine  $\rho_b$ in terms of the
various fields in the low-energy theory by solving the brane
junction conditions, including those of gravity.
\item It is this relaxation of $\rho_b$ that, in certain
circumstances, is ultimately responsible for the suppression   of
the contribution of the codimension-2 tension to the low-energy
on-brane curvature. In general, $\rho_b$ adjusts itself to ensure
that the effective potential, $U_2$, defined below the Kaluza
scale is completely determined by the brane tension, $T_2(\phi)$,
regarded as a function of the bulk scalar evaluated at the brane
position (given explicitly by eq.~\pref{constraintsoln}, of later
sections). In particular, as we will explain, the solution for
$U_2$ strictly vanishes when $T_2' = \exd T_2/\exd \phi = 0$, as
required by what is known about the back-reaction of codimension-2
pure tension branes. More generally, if $T_2'$ is nonzero but
small, then $U_2$ is suppressed by factors of order $\kappa^2
T_2'$, where $\kappa$ is the higher-dimensional Planck scale in
the bulk.
\end{itemize}

The above results arise in a calculation which evaluates loop
effects due to integrating out brane fields at the quantum level,
but only integrates out bulk fields (and in particular $\rho_b$)
within the bulk classical approximation. A crucial question
therefore asks how bulk loops might change the above picture. We
close the paper by taking a step in this direction by estimating
the contributions of the most dangerous (short-wavelength) bulk
loops within the more specific context where the bulk is
six-dimensional and supersymmetric. This extends earlier
calculations of the ultraviolet sensitivity of bulk loops far from
the brane, to include the effects of loops that are close to the
branes. We find that, for the contributions examined,
supersymmetry can suppress bulk loops to be of order the
Kaluza-Klein scale, again representing a significant suppression
to the low-energy potential $U_2$.

We organize our presentation as follows. \S2 starts by reviewing
the brane-bulk matching conditions for codimension-2 branes, as
recently derived in ref.~\cite{cod2ren}. This section in
particular describes how the codimension-2 brane can be
regularized in terms of a small codimension-1 brane, and relates
the properties of each to the other. \S3 then adds a massive field
to the brane and integrates it out at one loop, keeping track of
how this loop changes its interactions with the bulk fields. \S4
finally combines the results of the earlier sections, by
specializing them to the simple case where the brane-bulk
couplings are exponentials in the bulk scalar. The size of both
the codimension-2 brane tension, $T_2$, and the low-energy
effective scalar potential, $U_2$, are computed, both before and
after integrating out the massive brane field.
We conclude in \S5.

\section{The Framework}

We work for illustrative purposes within a higher-dimensional
scalar-tensor theory that provides the simplest context for
displaying our calculations. Since our interest is in integrating
out heavy matter on codimension-2 branes, we focus primarily on
the situations of space-filling $d$-dimensional branes sitting
within a $D=(d+2)$-dimensional bulk spacetime. The particular case
of $d=4$ and $D=6$ is of particular interest, as the simplest
`realistic' case within which the impact of higher-dimensional
ideas on technical naturalness might be relevant in practice.

\subsection{Bulk field equations}

Consider the following bulk action, governing the interactions
between the $D=(d+2)$-dimensional metric, $g_{\ssM\ssN}$ and a
real scalar field, $\phi$:\footnote{We use a `mostly plus'
signature metric and Weinberg's curvature conventions \cite{GnC}.}
\be \label{bulkaction}
    S_\ssB = - \frac{1}{2\kappa^2} \int \d^{D}x \sqrt{-g} \; \left[
    \, g^{\ssM\ssN} \Bigl( {\cal R}_{\ssM\ssN} + \partial_\ssM \phi \,
    \partial_\ssN \phi \Bigr)
    \right] + S_{\GH}\,,
\ee
where ${\cal R}_{\ssM\ssN}$ denotes the Ricci tensor built from
$g_{\ssM\ssN}$ and $S_{\GH} = \kappa^{-2} \int_{\partial M}
\exd^{D-1}x \sqrt{-\gamma} \; K$, denotes the Gibbons-Hawking
action \cite{GH}, which is required when using the Einstein field
equations in the presence of boundaries (as we do below). Here
$\gamma_{mn}$ denotes the induced metric on the boundary, and $K =
\gamma^{mn} K_{mn}$ is the trace of the boundary's extrinsic
curvature. (Since we are also interested in the case of
higher-dimensional supergravity, which also involve Maxwell and
Kalb-Ramond fields, and nontrivial scalar potentials, $V = V_0 \,
e^\phi$ \cite{NS}, in  section \S4 we discuss the extent to
which these features change our results.)

The corresponding field equations are
\be \label{bulkEOM}
    \Box \phi = 0 \quad \hbox{and} \quad
    {\cal R}_{\ssM\ssN} + \partial_\ssM \phi
    \, \partial_\ssN \phi = 0 \,.
\ee

In the immediate vicinity of a codimension-2 brane we imagine the
bulk fields to take an axially (transverse) and maximally
(on-brane) symmetric form
\ba \label{metricform}
    \d s^2 &=& \d \rho^2 + \hg_{mn} \, \d x^m \d x^n \nn\\
    &=& \d \rho^2 + e^{2 B} \, \d \theta^2
    + e^{2W} g_{\mu \nu}
    \, \d x^\mu \, \d x^\nu \,,
\ea
where $\rho$ denotes proper distance transverse to the brane,
$\theta \simeq \theta + 2\pi$ is the angular coordinate encircling
the brane, and the functions $B$, $W$ and $\phi$ are functions of
$\rho$ only. The on-brane metric, $g_{\mu\nu}$, is a
$d$-dimensional maximally symmetric Minkowski-signature metric
depending only on $x^\mu$.

\subsubsection*{Accidental bulk symmetries}

The field equations, eqs.~\pref{bulkEOM}, enjoy two accidental
symmetries, whose interplay with brane interactions will
be explored in
the following:

\begin{itemize}
\item {\it Axion symmetry:} The axion symmetry is defined by
\be \phi \to \phi + \zeta \,, \label{axionsym} \ee
for constant $\zeta$, with $g_{\ssM\ssN}$ held fixed.
\item {\it Scaling symmetry:} A scaling symmetry of the field
equations is
\be \label{scalesym}
 g_{\ssM\ssN} \to \lambda^2 g_{\ssM\ssN} \,,
\ee
with constant $\lambda$, and $\phi$ held fixed.
\end{itemize}

Both of these symmetries take solutions of the classical field
equations into distinct new solutions of the same equations, but
need not be respected by the couplings of the bulk fields to any
space-filling source branes, whose properties we next describe.

\subsection{Brane properties}

We imagine the bulk to be sprinkled with a number of space-filling
codimension-2 source branes, whose back-reaction dominates the
asymptotic near-brane behaviour of the bulk fields. In a
derivative expansion, their low-energy brane-bulk interactions are
governed by the action
\be
 S_b = - \int \d^dx \, \sqrt{- \gamma} \; \Bigl[ T_2(\phi) +
 X_2(\phi) \, \partial^\mu \phi \, \partial_\mu \phi
 + Y_2(\phi) \, R  + \cdots \Bigr] \,,
\ee
where $\gamma_{\mu\nu}$ denotes the induced metric on the brane
and the subscript `2' emphasizes that the brane has codimension 2
(by contrast with a codimension-1 branes to be considered
shortly). The ellipses represent further terms that arise at low
energies in a derivative expansion.

This brane action breaks the axion symmetry, eq.~\pref{axionsym},
if any of the coefficients, $T_2$, $X_2$ or $Y_2$, depend on
$\phi$. The tension term, $T_2$, also breaks the scaling symmetry,
eq.~\pref{scalesym}, unless $T_2$ is constant. The
higher-derivative terms always break the scaling symmetry, but can
preserve a diagonal combination of eqs.~\pref{axionsym} and
\pref{scalesym} corresponding to $\lambda^2 = e^{a\zeta}$ provided
$X_2, Y_2 \propto e^{a\phi}$.

At still lower energies the dynamics of the bulk-brane system is
normally dominated by very light modes, that are massless within
the purely classical approximation. These include the low-energy
$d$-dimensional metric, $g_{\mu\nu}$, possibly together with a
variety of moduli, $\varphi$, coming from $\phi$ or the metric
components. The dynamics of these modes below the Kaluza-Klein
(KK) scale is governed by a different effective $d$-dimensional
theory,
\be
 S_\eff = - \int \d^dx \, \sqrt{-g} \; \left[ U_\eff(\varphi)
 + \frac{1}{2\kappa_4^2} \, R + \cdots \right] \,,
\ee
obtained by integrating out all bulk KK modes as well as any heavy
brane states. At the purely classical level this action is
obtained by eliminating these states as functions of the light
fields using their classical equations of motion, and so depend on
the details of the classical bulk action.

In the classical approximation the contribution of the branes to
$S_\eff$ takes a  simple form. The accidental symmetries
guarantee the existence (classically) of a massless scalar mode
corresponding to shifts of $\phi$, so $\phi(\rho) = \varphi +
\delta \phi(\rho)$. The low-energy potential, $U_\eff$, turns out
to arise as a sum over local terms, each evaluated at the position
of a brane \cite{cod2ren}:\footnote{A similar result, summarized
in Appendix \ref{app:sugra}, holds less trivially for gauged,
chiral supergravity, despite the appearance there of a scalar
potential and nontrivial background fluxes \cite{BdRHT}.}
\be
 U_\eff(\varphi) = \frac1d\sum_b U_2[\phi(\rho_b; \varphi)] \,.
\ee
The fact that $U_2$ (defined more explicitly below) can vanish
even when $T_2 \ne 0$ is what allows codimension-2 branes having
nonzero tension to have flat on-brane geometries \cite{CG,SLED}.

\subsection{Brane-bulk matching}

It is the quantities $T_2(\phi)$ and $U_2(\phi)$ that dictate the
near-brane behaviour of bulk fields, through the matching
conditions.
For our purposes,
assuming the brane of interest to be situated at $\rho
= 0$, these become (see \cite{BdRHT,cod2ren}
for a complete discussion):
\ba \label{cod2match}
 \lim_{\rho \to 0} \left( e^{B + d W} \partial_\rho \phi
 \right) &=& \frac{\kappa^2 T_2'}{2\pi} \nn\\
 \lim_{\rho \to 0} \left( e^{B + d W} \partial_\rho W
 \right) &=& \frac{\kappa^2 U_2}{2\pi d} \\
 \lim_{\rho \to 0} \left( e^{B + d W} \partial_\rho B
 \right) &=& 1 - \frac{\kappa^2}{2\pi} \left[ T_2 + \left(
 \frac{d-1}{d} \right) U_2 \right] \nn\,.
\ea

\subsubsection*{Codimension-1 regularization}

A drawback of eqs.~\pref{cod2match} is the dependence of the
right-hand-side on quantities like $\phi(\rho=0)$, that need not
be well-defined if $\phi$ diverges as one approaches the brane
positions. This can be dealt with by defining an
alternative, renormalized, codimension-2 brane action,
as discussed in
\cite{cod2ren,GW,CdR} by elaborating on
 the work \cite{Georgi:2000ks}.
 On the other hand, for the aim of the present work,
 it is convenient to simply
regularize this divergence through the artifice of replacing the
codimension-2 brane with a very small cylindrical codimension-1
brane, situated at $\rho = \rho_b$ \cite{BdRHT,cod2ren,UVCaps},
with the interior geometry ($\rho < \rho_b$) capped off with a
smooth solution to the bulk field equations (see
fig.~\ref{fig:regcap}). We use capital latin indices, $x^\ssM$, to
describe all $D=d+2$ coordinates at once, reserving lower-case
indices, $x^m$, for coordinates on the $(d+1)$-dimensional
codimension-1 brane, and greek indices, $x^\mu$, for the $d$
codimension-2 brane directions.

The action on this codimension-1 brane is chosen to be
\be \label{Sregdef}
 S_\reg = - \int \d^{d+1}x \, \sqrt{-\gamma} \;
 \Bigl[ T_1(\phi) + Z_1(\phi) \, \partial^m \sigma
 \, \partial_m \sigma + \cdots \Bigr] \,,
\ee
where $\sigma$ is a massless, on-brane mode, whose presence is
included in order to dynamically support the brane radius at
nonzero $\rho$ against its propensity to collapse gravitationally.
This is done by choosing for its classical solution a
configuration that winds around the brane, $\sigma = n \theta$,
for $n$ a nonzero integer.

\EPSFIGURE[t]{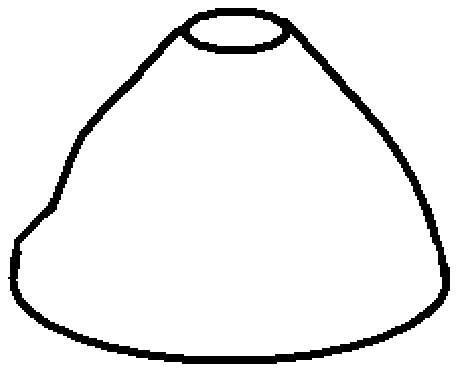, width = 0.4\textwidth,
height=5cm}{\sl The regularized near-brane cap
geometry.\label{fig:regcap}}

In terms of this action the codimension-2 tension is obtained
directly by dimensional reduction in the $\theta$ direction. Using
$e^{B(\rho_b)} = \rho_b$, this leads to
\be \label{T2def}
 T_2(\phi,\rho_b) = 2\pi \rho_b \left[ T_1(\phi) +
 \frac{n^2}{2\rho_b^2} \, Z_1(\phi) \right] \,.
\ee
As for the codimension-2 action, this preserves the axionic
symmetry, eq.~\pref{axionsym}, if $T_1$ and $Z_1$ are $\phi$
independent. However, because $\rho_b \to \lambda \rho_b$ there is
no choice for $T_1$ which preserves the scaling symmetry,
eq.~\pref{scalesym}. The diagonal combination with $\lambda =
e^{-a \zeta}$ survives if $T_1 \propto e^{a\phi}$ and $Z_1 \propto
e^{-a\phi}$.

The brane contribution, $U_2$, to the low-energy potential can
also be computed in terms of $T_1$ and $Z_1$ by classically
integrating out the bulk KK modes explicitly. This integration
involves evaluating the classical action at the classical
solution, with the result regarded as a function of the low-energy
zero modes, $g_{\mu\nu}$ and $\varphi$. Keeping in mind that the
only nonzero part of the action, eq.~\pref{bulkaction}, is in this
case the Gibbons-Hawking term, $S_\GH$, and that this receives
opposite-sign contributions from outside and inside the
codimension-1 brane, one obtains a result that depends only on the
jump conditions \cite{IJC} evaluated at the brane position
\ba \label{branebulkEFT}
    S_\eff(\varphi) = \left( S_{\ssB} + \sum_b S_\reg
    \right)_{\phi^{\rm cl}(\varphi)
    , g^{\rm cl}_{\ssM\ssN} (\varphi)}
    &=& \sum_b \left\{ S_\reg - \frac{1}{\kappa^2}
    \int \d^{d+1} x \, \sqrt{-\hg} \; \Bigl[ K \Bigr]_b \right\}
    \nn\\
    &=&  \sum_b \left\{ S_\reg - \frac{1}{d}
    \int \d^{d+1} x \, \sqrt{-\hg} \; \hg_{mn} T^{mn} \right\}
    \,,
\ea
where $\Bigl[ X \Bigr]_b = \lim_{\epsilon \to 0} \Bigl[ X(\rho_b +
\epsilon) - X(\rho_b - \epsilon) \Bigr]$. $\sqrt{-\hg} \; T^{mn} =
2 \, \delta S_\reg/\delta \hg_{mn}$ is the stress tensor of the
codimension-1 brane, whose independent components are
\be \label{T2Ttt}
 T_{\mu\nu} = - \left( T_1 + \frac{n^2}{2 \rho_b^2} \, Z_1
 \right) \hg_{\mu\nu} \quad \hbox{and} \quad
 T_{\theta\theta} = - \left( T_1 - \frac{n^2}{2 \rho_b^2} \, Z_1
 \right) \hg_{\theta\theta} \,.
\ee
Eq.~\pref{branebulkEFT} implies each brane contributes to the
very-low-energy theory as if its codimension-1 Lagrangian density
were
\be \label{hatLdef}
 \hat{\cal L}_\reg = {\cal L}_\reg - \frac2d \, \hg_{mn}
 \, \frac{\delta S_\reg}{\delta \hg_{mn}} \,.
\ee
Once compactified in the $\theta$ direction, this gives the
following brane contribution to $U_2$ \cite{cod2ren}:
\be \label{U2Tthth}
 U_2(\phi,\rho_b) = -2\pi \rho_b \left[ T_1(\phi) -
 \frac{n^2}{2\rho_b^2} \, Z_1(\phi) \right]
 = 2\pi \rho_b \, g^{\theta\theta} T_{\theta\theta} \,.
\ee

Finally, the expression for $\rho_b(\phi)$ is obtained by
integrating the ($\rho\rho$) Einstein equation, which expresses
the `Hamiltonian' constraint for integrating the bulk field
equations in the $\rho$ directions \cite{SN,BdRHT}:
\be \label{constraint}
 U_2 \left[ \frac{4\pi}{\kappa^2} - 2 T_2 - \left(
 \frac{d-1}{d} \right) U_2 \right] - \left( T_2' \right)^2
 \simeq 0 \,,
\ee
and the approximate equality involves dropping terms that are of
order $\rho_b^2 R$ relative to those displayed. Such curvature
terms are always negligible in the limit where the brane size,
$\rho_b$, is much smaller than the bulk radius of curvature to
which it gives rise.\footnote{Earlier authors often use this
constraint to determine $R$, but as argued in ref.~\cite{cod2ren},
this is not appropriate in the effective-field-theory limit, where
the brane is much smaller than the scales associated with the bulk
geometry.}

Eq.~\pref{constraint} states that the solution $\rho_b(\phi)$
adjusts itself to ensure that $U_2(\phi) = U_2(\phi,\rho_b(\phi))$
is not independent of $T_2(\phi) = T_2(\phi,\rho_b(\phi))$.
Expanding in powers
of $\kappa^2$, we find
\ba \label{constraintsoln}
 \frac{\kappa^2 U_2}{2\pi} &=& \left( \frac{d}{d-1} \right)
 \left\{ \left( 1 - \frac{\kappa^2 T_2}{2\pi} \right) - \left[
 \left( 1 - \frac{\kappa^2 T_2}{2\pi} \right)^2 -
 \left( \frac{d-1}{d} \right) \left( \frac{\kappa^2 T_2'}{2\pi}
 \right)^2 \right]^{1/2} \right\} \nn\\
 &\simeq& \frac12 \left( 1 - \frac{\kappa^2 T_2}{2\pi}
 \right)^{-1} \left( \frac{\kappa^2 T_2'}{2\pi}
 \right)^2 + \cdots \,.
\ea
The second line here emphasizes that the root is chosen to ensure
that $U_2$ vanishes in the limit when $T_2' \to 0$, since this
limit corresponds to rugby-ball type geometries \cite{CG,SLED}
having flat on-brane spacetimes ($U_2=0$) with nonzero but
$\phi$-independent tensions ($T_2' = 0$). It is simple
to see that all the corrections in higher powers
of $\kappa^2$, contained in the dots, are
proportional to $T_2'$.

Solving eq.~\pref{constraint} to lowest order in $\kappa^2$ leads
to the condition $U_2 \simeq 0$, and so
\be \label{rho0vsphi}
 \rho_b^2(\phi) \simeq \frac{n^2 Z_1(\phi)}{2 T_1(\phi)} \,,
\ee
showing that, in the limit in
which gravity is weak, $\rho_b$ adjusts itself to try
to set $U_2$ to zero. Using this solution, we can
integrate out the quantity $\rho_b$,
and  the codimension-2 brane
tension becomes
\be
 T_2(\phi) = T_2(\phi,\rho_b(\phi)) \simeq 2\pi |n|
 \sqrt{2 \, T_1(\phi) Z_1(\phi)} \,.
\ee
This expression then allows $U_2$ to be computed from
eq.~\pref{constraint} or \pref{constraintsoln} to next-to-leading
order in $\kappa^2$, giving \cite{cod2ren}
\be \label{U2vsT2}
 U_2(\phi) = U_2(\phi,\rho_b(\phi)) \simeq \frac{\kappa^2}{4\pi}
 \, \left( T_2' \right)^2 = \left( \frac{\pi n^2 \kappa^2}{2}
 \right) \frac{ \left[ \left( T_1 Z_1 \right)' \right]^2}{
 T_1 Z_1} \,.
\ee
As mentioned earlier, the function $U_2(\phi)$ vanishes when the
brane tension $T_2$ does not depend on the field $\phi$. Looking
at the first of equations (\ref{cod2match}), we see that having
$T_2$ be $\phi$-independent also requires the dilaton derivative
to vanish as one approaches the brane. Since such derivatives
naturally vanish when the brane is located in a region where
$\phi$ has a constant, or approximately constant, profile in the
bulk, it is natural to find that branes with $\phi$-independent
tensions are commonly the sources for geometries having such
regions. This observation will turn out to be useful in the
following.

\subsection{An Example}
\label{s:example}

For later purposes we pause here to record the above steps for an
interestingly broad example. We choose for this purpose the case
where $T_1$ and $Z_1$ are exponentials:
\be \label{expTZ}
 T_1 (\phi) = A_\ssT e^{- a_t \phi}
 \qquad \hbox{and} \qquad
 Z_1 (\phi) = A_\ssZ e^{-a_z \phi} \,,
\ee
and so
\be
 T_2(\phi,\rho_b) \simeq 2\pi \left[ \rho_b A_\ssT e^{-a_t \phi}
 + \left( \frac{n^2 A_\ssZ}{2 \rho_b} \right)
 e^{-a_z \phi} \right]\,,
\ee
and
\be
 U_2 (\phi,\rho_b) \simeq -2\pi \left[ \rho_b A_\ssT e^{-a_t \phi}
 - \left( \frac{n^2 A_\ssZ}{2 \rho_b} \right) e^{-a_z \phi}
 \right] \,.
\ee

In this case the zeroth-order brane size is
\be
 \rho_{b0} = |n| \sqrt{\frac{A_\ssZ}{2 A_\ssT}} \,
 e^{-(a_z- a_t)\phi/2} \,.
\ee
Using these,
 the leading contribution to the codimension-2 brane
tension and on-brane potential then become
\be
 T_2(\phi) \simeq {T_2}_0(\phi)
 = 2 \pi |n| \sqrt{2A_\ssT A_\ssZ}
 \, e^{-(a_t+ a_z) \phi/2} \,,
\ee
and
\be
 U_2 (\phi) \simeq
 \frac{\kappa^2}{4\pi} \left({{T_2}_0}' \right)^2
 = \frac{\pi}{2} \, n^2 (a_t+ a_z)^2 \kappa^2
 A_\ssT A_\ssZ e^{-(a_t+ a_z) \phi}
 \,.
\ee

Recall that these choices always break the scaling symmetry,
eq.~\pref{scalesym}, provided at least one of $A_\ssT$ or $A_\ssZ$
is nonzero. They respect the axionic symmetry,
eq.~\pref{axionsym}, if and only if $a_t = a_z = 0$. Finally, they
preserve a diagonal combination of these two symmetries if $a_t+
a_z = 0$. Notice that in this last
case $T_2$ is $\phi$-independent,
and so eq.~\pref{constraintsoln} shows $U_2 = 0$ solves the
constraint \pref{constraint} to all orders in $\kappa^2$.

\section{Integrating out a Massive Brane Field}

We next investigate the stability of the above considerations to
ultraviolet effects on the brane. The simplest way to do so is to
explicitly integrate out a heavy brane-localized field, and see
how the brane-bulk connection changes as a result.

\subsection{The brane field}

To this end, consider supplementing the brane action with new term
describing a massive real scalar field, $\psi$. Since our goal is
to see how this changes the bulk-brane interaction, we regard
$\psi$ as being localized on the codimension-1 regularized brane,
and so take $S_\reg \to S_\reg + S(\psi)$ with
\be
    S(\psi) = - \frac12 \int \d^{(d+1)}x \; \sqrt{-\hg} \Bigl[ P(\phi)
    (\partial \psi)^2 + m_\star^2 Q(\phi) \,\psi^2 \Bigr] \,.
\ee
Here $m_\star$ is a constant having dimensions of mass, and if we
adopt $z = \rho_b \theta$ as coordinate in the periodic direction,
then all fields satisfy the boundary condition $\psi(z) = \psi(z +
L)$, with $L = 2\pi \rho_b$.

For the purposes of computing quantum corrections, we imagine
starting with a classical solution whose induced metric on the
codimension-2 brane is flat, making the metric on the
codimension-1 regularizing brane
\be
    \exd s^2 = e^{2W_b} \eta_{\mu\nu} \,
    \exd x^\mu \exd x^\nu
    + \rho_b^2 \exd \theta^2 \,.
\ee
It is convenient at this point to re-scale $e^{W_b}$ into $x^\mu$,
so that $\exd s^2 = \eta_{mn} \, \exd x^m \exd x^n$ where we take
the coordinate in the angular direction to be $x^{d} = z$. As
discussed above, such a flat classical background would arise, for
instance, if $T_1 = A_\ssT e^{-a_t \phi}$ and $Z_1 = A_\ssZ
e^{-a_z \phi}$ with $a_z = - a_t$.

Provided $P \ne 0$ the $\psi$-particle action always breaks the
scaling symmetry, eq.~\pref{scalesym}, but preserves the axionic
symmetry if and only if $P$ and $Q$ are $\phi$-independent. A
diagonal subgroup of these two symmetries can be preserved when
$P$, $Q$, $T_1$ and $Z_1$ are all exponentials,
\be \label{expPQ}
 P (\phi) = A_\ssP \, e^{- a_p \, \phi} \quad \hbox{and} \quad
 Q (\phi) = A_\ssQ \, e^{- a_q \phi} \,,
\ee
provided $a_p = a_z = -a_q = -a_t$. The effective mass of $\psi$
for observers on the brane is $\phi$-dependent, given explicitly
by
\be \label{mvsphi}
    m^2(\phi) = \frac{m_\star^2 \, Q(\phi)}{P(\phi)} \,,
\ee
and so $m^2 = m_\star^2 (A_\ssQ/A_\ssP) e^{-(a_q-a_p)\phi}$ when
$P$ and $Q$ are exponentials, {\it \`a la} eqs.~\pref{expPQ}.

The stress energy for this heavy scalar is given by
\be
    T_{mn}(\psi) = P(\phi) \partial_m \psi \partial_n \psi
    - \frac12 \, g_{mn} \,
    \Bigl[ P(\phi) (\partial \psi)^2
    + m_\star^2 Q(\phi)
    \, \psi^2 \Bigr] \,,
\ee
which satisfies
\be
    {T^m}_m(\psi)
    = - \left(\frac{d-1}{2} \right) P(\phi)
    (\partial \psi)^2 - \left(
    \frac{d+1}{2} \right) \, m_\star^2 Q(\phi) \, \psi^2
\ee
in $d$ spacetime dimensions on the codimension-2 brane (with $d =
4$ being the case of most direct interest).

\subsection{Quantum Contributions}

To assess the contribution of quantum effects on the brane, we
integrate out $\psi$ by computing the Gaussian functional integral
\be \label{Gammadef}
    \exp \Bigl[ i \Gamma(\phi,g) \Bigr] = \int {\cal D}\psi \;
    \exp \Bigl[ i S(\psi,\phi,g) \Bigr] \,,
\ee
so that $\int {\cal D}\psi \; e^{i[S_\reg + S(\psi)]} =
e^{i[S_\reg + \Gamma]}$. Eq.~\pref{Gammadef} may be evaluated by
differentiating with respect to $\phi$, giving
\be \label{Gammaderiv}
    \frac{\delta \Gamma}{\delta \phi} =
    - \frac12 \sqrt{-\hg} \left[ \frac{P'}{P} \,
    \langle P \, (\partial \psi)^2 \rangle + \frac{Q'}{Q}
    \, \langle m_\star^2 Q \, \psi^2 \rangle
    \right] \,,
\ee
where $\langle X(\psi) \rangle = e^{-i \Gamma} \int {\cal D} \psi
\, X(\psi) \, e^{iS}$.

As described in Appendix \ref{app:EQC}, the relevant expectation
values can be expressed as
\be \label{dPsiSqExp}
    \langle P(\phi) \partial_\mu \psi \partial_\nu \psi \rangle
    =  \frac{\eta_{\mu\nu}}{2L^{d+1}}
    \int_0^\infty \frac{\d t}{t^{(d+2)/2}} \; e^{-\lambda t}
    \vartheta_3(it)
    = \eta_{\mu\nu} \, \frac{I_{1+d/2}(\lambda) }{2L^{d+1}}
    \,,
\ee
and
\be \label{dzPsiSqExp}
    \langle P(\phi) (\partial_z \psi)^2 \rangle =
    - \frac{i}{L^{d+1}} \int_0^\infty \frac{\d t}{t^{d/2}}
    \; e^{-\lambda t} \vartheta_3'(it)
    = - \frac{J_{d/2}(\lambda)}{L^{d+1}} \,,
\ee
and
\be \label{QPsiSqExp}
 \langle m_\star^2 Q(\phi) \, \psi^2 \rangle =
 \frac{\lambda}{L^{d+1}} \int_0^\infty
 \frac{\d t}{t^{d/2}} \; e^{-\lambda t} \vartheta_3(it)
 = \frac{\lambda I_{d/2}(\lambda) }{L^{d+1}} \,,
\ee
where
\be
 \lambda(\phi) = \frac{m^2(\phi) L^2}{4\pi} =
 \frac{m_\star^2 L^2 \, Q(\phi)}{4\pi P(\phi)}
 =  \pi \rho_b^2 m^2(\phi) \,.
\ee
The properties of the functions $I_\alpha(\lambda)$ and
$J_\alpha(\lambda)$ are spelt out in detail in Appendix
\ref{app:functions}. They satisfy a very useful identity,
\be \label{ExpIdentity}
    \Bigl\langle P(\phi) \,\partial_m \psi \, \partial^m \psi
    + m_\star^2 Q(\phi) \, \psi^2 \Bigr\rangle
    = \frac{1}{L^{d+1}} \left[ \frac d2 \, I_{1+d/2}(\lambda)
    - J_{d/2}(\lambda) + \lambda I_{d/2}(\lambda) \right] = 0 \,,
\ee
which (as is proven in Appendix \ref{app:EQC}) is a consequence of
our use of dimensional regularization.

Quantum fluctuations in $\psi$ contribute to the brane
stress-energy tensor, $T_{mn} \to T_{mn} + \Bigl \langle
T_{mn}(\psi) \Bigr \rangle$, where
\ba \label{TrEqn}
    \Bigl\langle T_{mn}(\psi) \Bigr\rangle &=& \frac{2}{\sqrt{-\hg}}
    \, \frac{\delta \Gamma}{\delta \hg^{mn}}
    = \Bigl\langle P(\phi) \, \partial_m \psi \, \partial_n \psi
    \Bigr\rangle - \frac12 \, \hg_{mn} \,
    \Bigl\langle P(\phi) \, (\partial \psi)^2
    + m_\star^2 Q(\phi) \, \psi^2 \Bigr\rangle \nn\\
    &=& \Bigl\langle P(\phi) \,\partial_m \psi \, \partial_n \psi
    \Bigr\rangle \,,
\ea
which uses the identity, eq.~\pref{ExpIdentity}. Its components
evaluate to
\ba
 \Bigl\langle T_{\mu\nu} (\psi) \Bigr\rangle
 &=& \eta_{\mu\nu} \; \frac{I_{1+d/2}(\lambda)}{2L^{d+1}}
 \nn\\
 \hbox{and} \quad
 \Bigl \langle T_{zz} (\psi) \Bigr \rangle
 &=& - \frac{J_{d/2}(\lambda)}{L^{d+1}}
 = - \frac{[\lambda I_{d/2}(\lambda) + \frac d2
 I_{1+d/2}(\lambda)]}{L^{d+1}} \,,
\ea
and, again using eq.~\pref{ExpIdentity}, its trace is
\ba \label{traceexp}
    \Bigl\langle {T^m}_m (\psi)\Bigr \rangle
    &=& -\left(\frac{d-1}{2} \right) \Bigl\langle
    P(\phi) \, (\partial \psi)^2 \Bigr \rangle -
    \left( \frac{d+1}{2} \right) \, \Bigl \langle m_\star^2 Q(\phi)
    \, \psi^2 \Bigr \rangle \nn\\
    &=&  \Bigl\langle
    P(\phi) \, (\partial \psi)^2 \Bigr \rangle
    = -\Bigl\langle m_\star^2 Q(\phi) \, \psi^2 \Bigr\rangle
    = - \frac{\lambda I_{d/2}(\lambda)}{L^{d+1}}\,,
\ea
which (naively) vanishes when $m_\star=0$, and so is completely
given by any trace anomaly when this is nonzero. In particular,
notice that $m_\star = 0$ implies $\langle {T^m}_m \rangle = 0$
when $d = 2k$ is a positive even integer, since in this case the
codimension-1 brane has odd dimension and so the divergent parts
of the above expressions vanish in dimensional regularization (see
Appendix \ref{app:EQC} for details).

\subsection{Codimension-2 quantities}

For the present purposes, of most interest is the contribution of
$\psi$ loops to low energy quantities, so we next seek the loop
contribution to the codimension-2 quantities $T_2$ and $U_2$. This
involves repeating their earlier derivation with the replacements
$S_\reg \to S_\reg + \Gamma$ and $T_{mn} \to T_{mn} + \langle
T_{mn}(\psi) \rangle$.

The most direct means for computing both $T_2$ and $U_2$ then uses
their representation as compactifications of components of the
stress tensor --- $T_2 = -L \, g^{tt} T_{tt}$ and $U_2 = L \,
g^{\theta \theta} T_{\theta \theta}$ --- which summarize
eqs.~\pref{T2def}, \pref{T2Ttt} and \pref{U2Tthth}. These remain
true provided $T_{mn} \to T_{mn} + \langle T_{mn}(\psi) \rangle$,
suggesting that the change generated by $\psi$ loops is
\be \label{DT2}
 \Delta T_2(\phi,L) = L \langle T_{tt} \rangle =
 - \frac{I_{1+d/2}(\lambda)}{2L^d} \,,
\ee
and
\be \label{DU2}
 \Delta U_2(\phi,L) = L \langle T_{zz} \rangle =
 - \frac{J_{d/2}(\lambda)}{L^d} = - \frac{[\lambda
 I_{d/2}(\lambda) + \frac d2 I_{1+d/2}(\lambda)]}{L^d}
 \,.
\ee
Notice in particular that the $\phi$-dependence of these
quantities (at fixed $L$) only enters through the combination
$m^2(\phi) \propto Q/P$.

A check on these expressions comes if we instead work directly
with the loop contributions to the regularized brane action,
$S_\reg = \int \d^{d+1}x \, \sqrt{-\hg} \; {\cal L}_\reg$. Writing
$S_\reg + \Gamma = \int \d^{d+1}x \, \sqrt{-\hg} \; ( {\cal
L}_\reg + \Delta {\cal L}_\reg)$, eq.~\pref{Gammaderiv} implies
\be \label{Lregderiv}
    \frac{\partial \Delta {\cal L}_\reg}{\partial \phi}
    = \frac12 \left[ \frac{P'}{P} - \frac{Q'}{Q}
    \right] \langle m_\star^2 Q \, \psi^2 \rangle
    = - \frac{1}{2m^2} \left( \frac{\partial m^2}{
    \partial \phi} \right) \frac{\lambda I_{d/2}(\lambda)
    }{L^{d+1}} \,,
\ee
which uses $(P'/P) - (Q'/Q) = [\ln(P/Q)]' = -(1/m^2)(\partial
m^2/\partial\phi)$, eq.~\pref{QPsiSqExp} and
eq.~\pref{ExpIdentity}. Finally, provided $\partial m^2/\partial
\phi \ne 0$ this integrates to
\be \label{DLreg}
 \Delta {\cal L}_\reg = -\frac{1}{2 L^{d+1}}
  \int^\lambda \d \hat\lambda \,  I_{d/2}(\hat\lambda)
  = \frac{I_{1+d/2}(\lambda)}{2L^{d+1}} \,,
\ee
where the integral is performed using $\int^\lambda \d\hat\lambda
\, I_\alpha (\hat\lambda) = -  I_{1+\alpha}(\lambda)$. Finally,
dimensional reduction gives $\Delta T_2 = - L \, \Delta {\cal
L}_\reg$, in agreement with eq.~\pref{DT2}.

\EPSFIGURE[t]{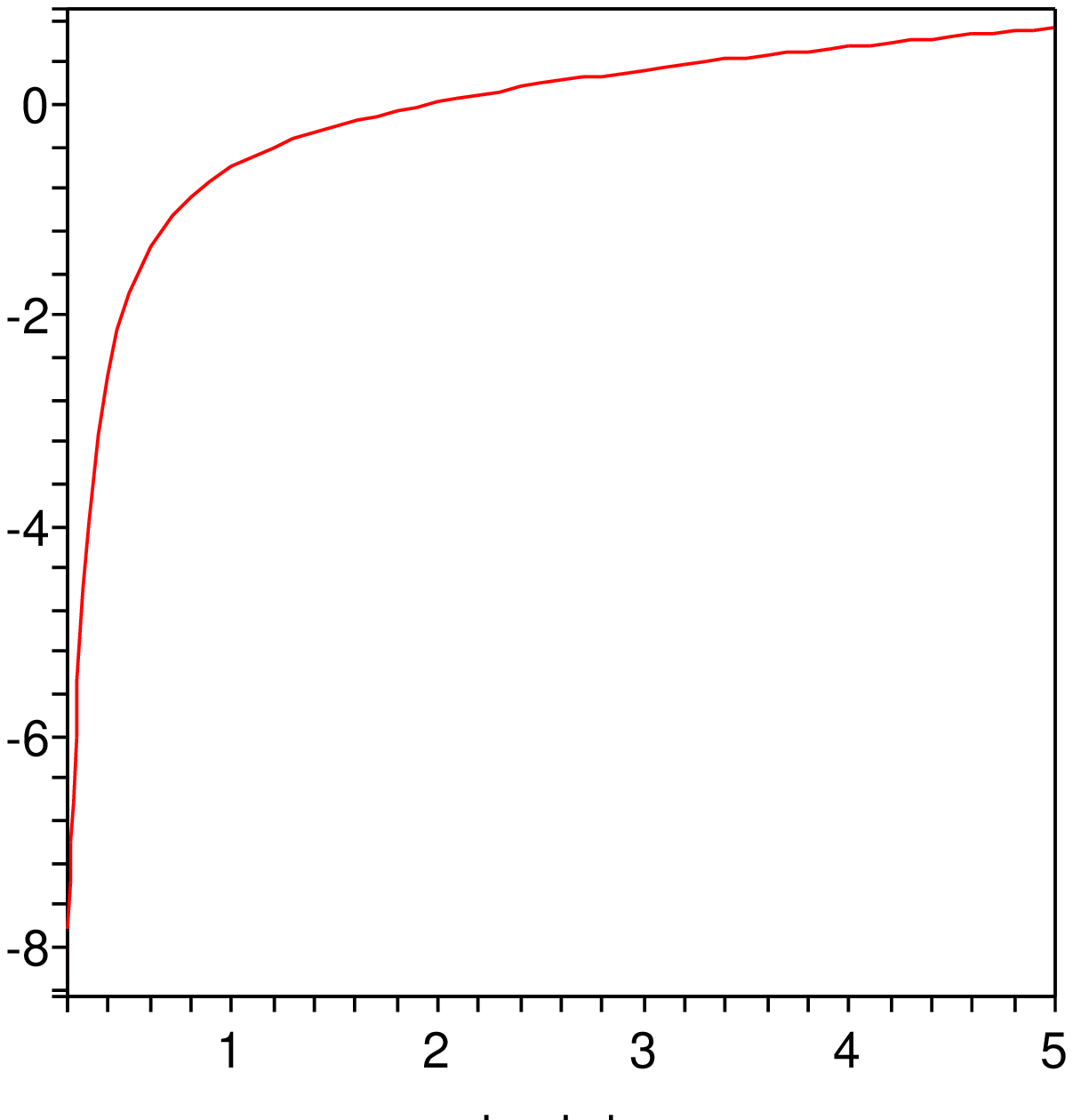, width = 0.5\textwidth, height=7cm}{\sl The
function $U_2(\lambda) + \Delta U_2(\lambda)$ vs
$\lambda$.\label{fig:U2plot}}

Similarly, integrating out the bulk KK modes at the classical
level using $S_\reg + \Gamma$ as the regularized brane action,
leads to loop corrections to the effective action for energies
well below the transverse KK scale, as seen by observers residing
on the brane. The change to these due to the $\psi$ loops is
\be
    \Delta S_{\rm eff} = \Gamma -
    \frac1d \int \d^{d+1}x \, \sqrt{-\hg} \;
    \langle {T^m}_m \rangle  \,,
\ee
showing that $\psi$ loops enter into this action both by directly
changing the brane action, and by changing the junction conditions
relevant to integrating out the bulk fields. Writing $\Delta
S_\eff = \int \exd^{d}x \sqrt{-g} \; \Delta {\cal L}_\eff$ gives
$\Delta {\cal L}_\eff = \int \exd z \left[\Delta {\cal L}_\reg -
\frac1d \, \sqrt{-\hg} \; \langle {T^m}_m \rangle \right]$, and so
\be \label{DLeff}
 \Delta {\cal L}_\eff = \frac{1}{L^{d}} \left[ \frac12 \,
 I_{1+d/2}(\lambda) + \frac{\lambda}{d} \,
 I_{d/2}(\lambda) \right]
 = \frac{J_{d/2}(\lambda)}{d L^{d}} \,,
\ee
so using $\Delta U_2 = - d\, \Delta {\cal L}_\eff$ agrees with
eq.~\pref{DU2}.

Finally, notice that the derivation of the curvature constraint,
eq.~\pref{constraint}, relating $U_2$ to $T_2$, goes through as
before, but with $T_{mn}$ replaced everywhere by $T_{mn} + \langle
T_{mn}(\psi) \rangle$.  This guarantees that $U_2 + \Delta U_2$
can be obtained from $T_2 + \Delta T_2$ by using
eq.~\pref{U2vsT2}, precisely as $U_2$ can from $T_2$.

\subsubsection*{The limit $m L \gg 1$}

Before going further it is instructive to evaluate these for the
asymptotic case where $mL \gg 1$, since this is particularly easy
to interpret. Appendix \ref{app:functions} shows that when
$\lambda \gg 1$ we have $I_\alpha(\lambda) \simeq \lambda^{\alpha
- 1/2} \Gamma \left( \frac12 - \alpha \right)$, and so
\be
 \Delta T_2 \simeq  - \frac{m^{d+1}L}{2(4\pi)^{(d+1)/2}} \; \Gamma \left(
 - \frac12 - \frac d2 \right)
 \,, \ee
and
\be
 \Delta U_2 \simeq \frac{m^{d+1}L}{2(4\pi)^{(d+1)/2}} \;
 \Gamma \left( -\frac12 - \frac d2 \right)
 \,.
\ee

Notice these satisfy $\Delta U_2 = - \Delta T_2$, just as would be
expected if both arose from a contribution to the codimension-1
brane tension,
\ba \label{T1mlarge}
 \Delta T_1(\phi) &\simeq&
 - \frac{m^{d+1}}{2(4\pi)^{(d+1)/2}} \;
 \Gamma \left( -\frac12 - \frac d2 \right) \nn\\
 &\to&  \frac{m^5}{120 \pi^2} \quad \hbox{as} \quad
 d \to 4 \,,
\ea
where the limit $d \to 4$ uses $\Gamma\left( - \frac52 \right) =
-\frac{8}{15} \, \sqrt\pi$. Indeed, when $\psi$ is very massive
compared with the codimension-1 KK scale, $2\pi/L = 1/\rho_b$, we
expect its quantum effects are well captured by local
contributions to the brane action, starting with $\Delta {\cal
L}_\reg \simeq - \Delta T_1 +$ (higher derivative terms).
Furthermore, eq.~\pref{T1mlarge} agrees with the large-$m$ limit
of eq.~\pref{DLreg}.

In the general case eqs.~\pref{DT2} and \pref{DU2} predict $\Delta
T_2 \ne - \Delta U_2$, implying they do not have an interpretation
as simple as a contribution to $T_1$.\footnote{This is a special
case of the more general observation that the radius dependence of
Casimir energies on torii \cite{CETori} cannot be represented in
terms of local curvature invariants.} This is because when $m L
\lsim 1$ the wavelengths integrated out are similar to the size of
the entire circular direction on the codimension-1 brane, and so
need not depend only on local geometrical quantities. Their
contributions must be local in the transverse dimensions, however,
provided that their wavelength is much smaller than the typical
scales set by the transverse geometry.

\subsubsection*{The limit $mL \to 0$}

A second instructive limit corresponds to taking $mL \to 0$, for
which Appendix \ref{app:functions} shows $I_\alpha (0) = 2
\pi^{\frac12 - \alpha} \, \Gamma\left( \alpha - \frac12 \right) \,
\zeta \left(2\alpha - 1 \right)$. In this case eqs.~\pref{DT2},
\pref{DU2} and \pref{DLreg} become
\ba \label{DT20}
 \Delta T_2(\phi,L) \simeq \frac{\Delta U_2(\phi,L)}{d}
 \simeq - L \, \Delta {\cal L}_\reg
 &\simeq& - \frac{1}{\pi^{(d+1)/2} L^d} \, \Gamma \left(
 \frac12 + \frac d2 \right)
 \, \zeta \left( d+1 \right) \nn\\
 &\to& - \frac{3 \, \zeta \left( 5 \right)}{4\pi^{2} L^4}
 \quad \hbox{as} \quad d \to 4
 \,.
\ea
Notice $d \Delta T_2 \to \Delta U_2$ when $mL \to 0$, consistent
with the result $\langle {T^m}_m \rangle \to 0$.

\subsubsection*{Eliminating $\rho_b$}

After determining the expressions for $\Delta T_2(\phi, L)$
and $\Delta U_2(\phi, L)$,
the final step amount to eliminate  $L$  --- or equivalently
$\rho_b$
 --- in terms of $\phi$ by using
eq.~\pref{constraint}. To leading order in $\kappa^2$ this
involves solving the condition $U_2(\phi,L) \simeq 0$, leading to
eq.~\pref{rho0vsphi}, $\rho^2_{b0} \simeq n^2 Z_1/(2\, T_1)$ in
the absence of the quantum $\psi$ contributions.

For the purposes of computing the quantum correction $\delta
\rho_b^2$, suppose $T_1 \simeq M^{d+1}$ and $Z_1 \simeq M^{d-1}$
are characterized by a common regularization mass scale $M$, and
so eq.~\pref{rho0vsphi} implies $L_0 = 2\pi \rho_{b0} \simeq 2\pi
n/M$. Since for us $\rho_b$ plays the role of an ultraviolet
regulator, our interest in what follows is in the limit $m L_0 =
2\pi \rho_{b0} m \simeq 2 \pi nm/M \ll 1$ and $m/M \ll 1$, in
which case $d\, \Delta T_2(\phi,\rho_b) \simeq \Delta
U_2(\phi,\rho_b) \simeq -c\,d/(2\pi \rho_b)^d$ with $c =
\Gamma\left(\frac12 + \frac d2 \right) \zeta(d+1) \,
\pi^{-(d+1)/2}$. Since in this limit we have
\be
 L_0 T_1 \sim L_0 \left( \frac{n^2 Z_1}{\rho_{b0}^2}
 \right) \sim 2\pi n M^d \quad \hbox{and} \quad
 \Delta U_2 \sim \frac{d}{(2 \pi L_0^2)^{d/2}}
 \sim \frac{d M^d}{(2\pi)^{3d/2} n^d} \,,
\ee
it follows that $\left| \delta \rho_b^2 \right| \ll \rho_{b0}^2$,
provided
that the inequality
$(2\pi)^{1+3d/2} n^{d+1} \gg d$
is satisfied.
  Using this observation,
we solve $U_2 (\phi,\rho_b) \simeq 0$ perturbatively in $\delta
\rho_b^2$, to give
\be
 \frac{\delta \rho_b^2}{\rho_{b0}^2}
 \simeq -\frac{c \, d}{2 \pi^2 n^2 Z_1(\phi)
 (2\pi \rho_{b0})^{d-1}} = - \frac{c \, d}{T_1}
 \left( \frac{T_1}{2 \pi^2 Z_1} \right)^{(d+1)/2}
 \,.
\ee

With this choice, the leading corrections to $T_2(\phi)$ become
\ba \label{DT2QC}
 T_2(\phi) &=& {T_2}_0[\phi, \rho_{b0}(\phi)]
 + \frac{\partial {T_2}_0}{\partial \rho_b}
 [\phi, \rho_{b0}(\phi)] \, \delta\rho_b +
 \Delta T_2[ \phi, \rho_{b0}(\phi)] + \cdots \nn\\
 &=& 2\pi|n| \sqrt{2\, T_1(\phi) Z_1(\phi)}
 - \frac{c}{(2\pi |n|)^d} \left[ \frac{2\, T_1(\phi)}{Z_1(\phi)}
 \right]^{d/2} + \cdots \,,
\ea
which uses ${T_2}_0(\phi,\rho_b) = 2\pi \rho_b T_1 +\pi n^2
Z_1/\rho_b$ and so $\partial {T_2}_0/\partial \rho_b \propto
{U_2}_0(\phi,\rho_b)$ vanishes when evaluated at $\rho_b =
\rho_{b0}(\phi)$. In the previous expression, the dots
contain corrections proportional to positive powers of
$\kappa^2$.

Finally, the constraint, eq.~\pref{constraint}, automatically
ensures $U_2(\phi)$ satisfies eq.~\pref{constraintsoln}, and so
\be
 U_2(\phi) \simeq \frac12 \left( 1 - \frac{\kappa^2 T_2}{
 2\pi} \right)^{-1} \left( \frac{\kappa^2 T_2'}{2\pi}
 \right)^2 + \cdots \,,
\ee
up to higher powers of $(\kappa^2 T_2')^2$, with $T_2(\phi)$ given
by eq.~\pref{DT2QC}.

\section{Technical Naturalness}

We may now use the above tools to quantify how integrating heavy
brane physics modifies properties of the low-energy world. We ask
in particular how symmetry-breaking effects on the brane modify at
low energies the symmetries of the bulk action. We use for this
purpose both the scaling and axionic shift symmetries,
eqs.~\pref{axionsym} and \pref{scalesym}, of the simplified bulk
theory used for illustrative purposes here, but we have in mind
applications to other symmetries like supersymmetry as well.

\subsection{Brane Loops}

With this in mind we work within the particularly interesting
framework of exponential brane couplings to $\phi$, as described
by eqs.~\pref{expTZ} and \pref{expPQ} above: $T_1 = A_\ssT e^{-a_t
\phi}$, $Z_1 = A_\ssZ e^{-a_z \phi}$, $P = A_\ssP e^{-a_p \phi}$
and $Q = A_\ssQ e^{-a_q \phi}$. In this case the leading
contributions to $T_2(\phi)$ and $\rho_b(\phi)$ become
\ba \label{expT20}
 {T_2}_0(\phi) &=& 2\pi |n| \sqrt{2 A_\ssT A_\ssZ} \,
 e^{-(a_t + a_z) \phi/2} \nn\\
 \hbox{and} \quad
 \rho_{b0}(\phi) &=& |n| \sqrt{ \frac{A_\ssZ}{2 A_\ssT}}
 \, e^{-(a_z - a_t) \phi/2} \,,
\ea
and so ${U_2}_0 \simeq \kappa^2 (T_2')^2/4\pi$ becomes
\be \label{expU20}
 {U_2}_0(\phi) \simeq \frac{\pi n^2}{2} \,\kappa^2
 (a_t + a_z)^2 A_\ssT A_\ssZ
 \; e^{-(a_t + a_z)\phi} \,.
\ee
The $\psi$ mass, $m(\phi)$, is
\be \label{expmass}
 m(\phi) = m_\star \sqrt{\frac{A_\ssQ}{A_\ssP}} \,
 e^{-(a_q - a_p) \phi/2} \,,
\ee
and so
\be \label{explambda}
 \lambda_0 = \pi \rho_{b0}^2 m^2
 \simeq \frac{\pi n^2 m_\star^2}{2}
 \left( \frac{A_\ssZ A_\ssQ}{A_\ssT A_\ssP} \right)
 \, e^{-(a_z + a_q - a_t - a_p)\phi} \,.
\ee
Provided $\lambda_0 \ll 1$ the leading correction to the
codimension-2 tension due to $\psi$ loops, eq.~\pref{DT2QC},
becomes
\be \label{expDT2}
 \Delta T_2(\phi) \simeq - \frac{c}{(2\pi |n|)^d}
 \left[ \frac{2 A_\ssT}{A_\ssZ}
 \right]^{d/2} e^{-d(a_t - a_z) \phi/2}
\ee
and so
\be \label{expDU2}
 \Delta U_2(\phi) \simeq \frac{\kappa^2}{4\pi} \,
 \left[ \Delta T_2'(\phi) \right]^2
 \simeq \frac{\kappa^2 c^2d^2(a_t - a_z)^2}{16\pi
 (2\pi |n|)^{2d}}
 \left[ \frac{2 A_\ssT}{A_\ssZ}
 \right]^{d} e^{-d(a_t - a_z) \phi}
\ee

We consider now several important special cases.

\subsubsection*{The case $a_t = a_q := a$ and $a_z = a_p := a + 2\,b$}

This choice is motivated by a situation of practical interest
where there exists a frame for which $\phi$ appears
undifferentiated only as an overall power of $e^\phi$
pre-multiplying the entire brane action. That is, ${\cal
L}_\reg(\phi,\partial_m \phi,g_{mn}) = e^{-\alpha \phi}
f(\partial_m \phi, \cg_{mn})$ for some metric $\cg_{mn} =
e^{-2\beta \phi} g_{mn}$, implying $a_t = a_q = \alpha + d\,
\beta$ and $a_z = a_p = \alpha + (d-2) \beta$, so $a = \alpha +
d\, \beta$ and $b = -\beta$.

In this case we have $\rho_{b0} \propto e^{-b\phi}$ and $m \propto
e^{b\phi}$ so $\lambda_0 \propto m^2 \rho_{b0}^2$ is
$\phi$-independent. Then ${T_2}_0 \propto e^{-(a + b)\phi}$, and
the leading quantum correction $\Delta T_2 \propto e^{db \phi}$
(regardless of whether $\lambda_0$ is large or small). There are
then two special situations of particular interest:

\begin{enumerate}
\item {\it The case $b=0$}: This situation corresponds to it being
the bulk Einstein frame for which the brane action has the form
${\cal L}_\reg(\phi,\partial_m \phi,g_{mn}) = e^{-\alpha \phi}
f(\partial_m \phi, g_{mn})$. In this case, if $A_\ssT \sim
M^{d+1}$ and $A_\ssZ \sim M^{d-1}$ then we have ${T_2}_0 \propto
M^d e^{-a\phi}$ while $\Delta T_2 \propto M^d/(2\pi n)^{2d}$ is
precisely $\phi$-independent. Consequently in this case we have
${U_2}_0 \sim \kappa^2 M^{2d} a^2 e^{-2a\phi}$, while $\Delta U_2$
precisely vanishes to leading order in $\kappa^2$.
\item {\it The case $b = -a$}: In this situation $a_t = a_q = -
a_z = - a_p$, which is the condition that the brane action
preserves a diagonal combination of the two bulk symmetries,
eqs.~\pref{axionsym} and \pref{scalesym}. In this case it is
${T_2}_0(\phi)$ that is $\phi$-independent and so ${U_2}_0 \simeq
0$, while $\Delta T_2 \propto M^d e^{-da \phi}/(2\pi n)^{d}$ and
so $\Delta U_2 \sim \kappa^2 a^2 d^2 M^{2d} e^{-2da\phi}/(2\pi
n)^{2d}$. In this case the leading contribution to $U_2$ first
arises suppressed both by a loop factor and a power of $\kappa^2
M^d$.
\end{enumerate}

The virtues of both of these last two choices are combined in the
most remarkable situation: the case $a = b = 0$ (or $a_t = a_z =
a_p = a_q = 0$). This is the simplest choice, for which the brane
does not couple at all to the bulk scalar $\phi$, such as might be
required if the brane couplings must preserve the bulk shift
symmetry, eq.~\pref{axionsym}. It also captures the case of pure
gravity, for which there is no scalar field in the bulk to which
to couple. Such branes are known to arise in geometries having a
region where the scalar profile is constant in the
extra-dimensions.

In this case physical scales on the brane, like $\rho_b$ and $m$,
are $\phi$-independent, as is $T_2$. And this $\phi$-independence
holds provided only that shift-symmetry breaking effects (like
anomalies) are negligible. But because $T_2'$ vanishes so
robustly, eq.~\pref{constraintsoln} shows that the same is true
for $U_2$, which vanishes to all orders in $\kappa$ (within the
approximation of a classical bulk). This states that the brane
tension does not contribute at all towards the low-energy
potential, $U_\eff$, governing the on-brane curvature, much as the
geometry produced by a cosmic string is locally flat in 4
dimensions \cite{cosmicstrings}. Furthermore, on-brane loops do
not alter this property provided that they also do not couple to
$\phi$.

\subsection{Bulk Loops and Supersymmetry}

A drawback of the preceding discussion is its omission of quantum
effects in the bulk. In general these loops can be problematic,
particularly if they break the symmetries of interest. Although
this can be controlled for axionic symmetries, this need not be so
for features of the low-energy action, that are consequences of the
bulk scaling symmetry. It is here that supersymmetry in the bulk
can play a helpful role.

\subsubsection*{Changes due to a supersymmetric bulk}

Supersymmetry changes the above analysis in several important
ways, which we briefly summarize in this section (see Appendix
\ref{app:sugra} for more details). To keep things concrete we
focus on how the above discussion changes if the bulk is described
by chiral, gauged supergravity in six dimensions \cite{NS},
although many features generalize to other higher-dimensional
supergravities. In this case the action describing the classical
dynamics of the bosonic degrees of freedom has the form
\ba \label{6DSugraAction}
    \frac{{\cal L}_\ssB}{\sqrt{-g}} &=& - \frac{1}{2 \kappa^2} \,
    g^{\ssM\ssN} \Bigl[ R_{\ssM\ssN} + \partial_\ssM \phi \,
    \partial_\ssN \phi \Bigr] - \frac{2g^2}{\kappa^4} \; e^\phi \nn\\
    && \qquad\qquad - \frac14 \, e^{-\phi} \, F^a_{\ssM\ssN} F_a^{\ssM\ssN}
    - \frac{1}{2 \cdot 3!} \, e^{-2\phi} \, G_{\ssM\ssN\ssP}
    G^{\ssM\ssN\ssP} \,,
\ea
where $\phi$ is the 6D scalar dilaton, $G_{\ssM\ssN\ssP}$, is the
field strength for a Kalb-Ramond potential, $B_{\ssM\ssN}$,
arising in the gravity supermultiplet and $F^a_{\ssM\ssN}$ is the
field strength for the potential, $A^a_\ssM$, appearing in a gauge
supermultiplet. The parameter $g$ is the gauge coupling for a
specific gauge group, and has dimensions of inverse mass. Matter
scalars, $\Phi^i$, could also appear, but these are set to zero in
the above action, as is consistent with their field equations.

An important feature of this system is the existence of many
explicit solutions to the field equations describing
compactifications of two of the dimensions whose size is supported
by extra-dimensional gauge fluxes, $F^a_{\ssM\ssN}$
\cite{SS,SLED,6Dsolns}. For the simplest of these the compact
geometry is that of a 2-sphere, whose radius is fixed in terms of
$\phi$ by the equations of motion to satisfy \cite{SS,SLED}
\be \label{E:phircond}
 {r}^2 = {\kappa^2 e^{-\phi} \over 4 g^2} \, .
\ee

The value of $\phi$ itself is not fixed, despite the presence of a
nontrivial scalar potential. Its undetermined value represents a
classically flat direction, whose presence may be understood as a
consequence of a scale invariance having the form of a diagonal
combination of eqs.~\pref{axionsym} and \pref{scalesym}: $e^\phi
\to \omega \, e^\phi$ and $g_{mn} \to \omega^{-1} g_{mn}$. The
existence of this classical symmetry is most easily seen from the
existence of a frame for which the action has the form ${\cal
L}_\ssB = e^{-2\phi} {\cal F}(\partial_\ssM \phi,
\cg_{\ssM\ssN})$, where $\cg_{\ssM\ssN} = e^\phi g_{\ssM\ssN}$,
since in this frame the symmetry corresponds to shifting $\phi$
with $\cg_{\ssM\ssN}$ held fixed.

Almost all of the compactifications of this theory to 4 dimensions
involve singularities in the extra-dimensional geometry, which can
be interpreted as the singularities due to the back-reaction of
space-filling codimension-2 source branes situated about the bulk.
The low-energy action for these branes can be worked out using the
same trick used here of a regularizing codimension-1 brane, with
the complication that the Maxwell fields, $A^a_\ssM$, must also
satisfy junction conditions at the branes as well as Dirac
quantization conditions \cite{UVCaps,BdRHT}. In this case the
brane respects the scale invariance of the bulk classical field
equations only when $T_1 = A_\ssT e^{\phi/2}$ and $Z_1 = A_\ssZ
e^{-\phi/2}$. Another bulk symmetry that is sensitive to the
presence of branes is  supersymmetry. Singular sources, as
codimension two branes, in general break supersymmetry in the
bulk. The only exception are branes embedded in configurations
with constant bulk scalar, and that couple in a very specific way
to bulk fields. We will not elaborate on this interesting topic
(see  \cite{HyunMin} for detailed discussions), but we will return
to discuss the effects of brane supersymmetry breaking at the end
of this section.

\smallskip

The contributions of the branes to the very-low-energy action,
${\cal L}_\eff$, can be evaluated much as was done above by
integrating out the bulk KK modes at the classical level, and this
again leads to a remarkably simple result \cite{BdRHT} involving
only quantities localized at the branes. There turns out to be a
new contribution to $U_\eff$, however, because the bulk action,
eq.~\pref{6DSugraAction}, does not give zero when evaluated at a
classical solution. Instead, use of the Einstein and $\phi$ field
equations shows that
\be
 {\cal L}_\ssB(g^{cl}_{\ssM\ssN},\phi^{cl},\cdots)
 = \frac{1}{2 \, \kappa^2} \, \Box \, \phi^{cl} \,,
\ee
which gives a contribution proportional to the jump
$[\partial_\rho \phi ]_b$ across the position of each regularized
brane. The result is that the brane contribution to $U_\eff$ may
be computed by dimensional reduction, as if the regularized
codimension-1 brane action is given (since $d=4$) by \cite{BdRHT}
\be
 \hat {\cal L}_\reg = {\cal L}_\reg - \frac12 \, \hg_{mn}
 \frac{\partial \L_\reg}{\partial \hg_{mn}} - \frac12
 \frac{\partial {\cal L}_\reg}{\partial \phi}
 \,,
\ee
rather than eq.~\pref{hatLdef}. Consequently
\be
  U_\eff = \sum_b \left( \frac{U_2}{4} - \frac{T_2'}{2} \right)
  \simeq - \sum_b \frac{T_2'}{2} \Bigl[ 1 +
  {\cal O}\left( \kappa^2 T_2' \right) \Bigr]   \,,
\ee
where the approximate equality neglects $\kappa^2 (T_2')^2$
relative to $T_2'$. An identical expression holds for $\Delta
U_\eff$ in terms of $\Delta U_2$ and $\Delta T_2'$.

For instance, in the case considered above, with $T_1 = A_\ssT
e^{-a_t \phi}$, $Z_1 = A_\ssZ e^{-a_z \phi}$, $P = A_\ssP e^{-a_p
\phi}$ and $Q = A_\ssQ e^{-a_q \phi}$, eqs.~\pref{expT20} through
\pref{expDU2} for ${T_2}_0$, ${U_2}_0$, $\Delta T_2$ and $\Delta
U_2$ remain unchanged (but with $d=4$), while the low-energy
potential becomes $U_\eff \simeq -\frac12 \, T_2'$, so
\be
 {U_\eff}_0 \simeq \pi |n| (a_t + a_z) \sqrt{ \frac{A_\ssT A_\ssZ}
 {2}}
 \; e^{-(a_t + a_z) \phi/2} \,,
\ee
and
\be \label{expDT2new}
 \Delta U_\eff \simeq - \frac{c (a_t - a_z)}{(2\pi |n|)^4}
 \left[ \frac{2 A_\ssT}{A_\ssZ}
 \right]^{2} e^{-2(a_t - a_z) \phi} \,,
\ee
which assumes $m \rho_b \ll 1$.

Again ${T_2}_0'$ and ${U_\eff}_0$ both vanish in the
scale-invariant case where $a_z = - a_t = \frac12$, in which case
the dominant loop-generated correction becomes
\be
 \Delta U_\eff \simeq c\left( \frac{M e^{\phi/2}}{2\pi |n|}
 \right)^4 = c\left( \frac{\kappa M}{4\pi |n| g r} \right)^4 \,,
\ee
where $M^2 = 2 A_\ssT/A_\ssZ$ and we use the bulk field equation
$e^\phi = \kappa^2/(4g^2r^2)$ ({\it i.e.} eq.~\pref{E:phircond}).
Notice that this is both positive and of order $1/r^4$ when
$\kappa$, $g$ and $M$ are all of order the TeV scale --- with
$\kappa M/g \sim 0.1$, say, to allow the semiclassical
approximations used --- and so can be much smaller than the TeV
scale.

Of course such a small contribution to $U_\eff$ is not so
impressive unless it is much smaller than the physical mass of the
particle that was integrated out, and this is not generically so.
For example, in the scale-invariant case we have $\rho_b \simeq
|n|\sqrt{A_\ssZ/(2 A_\ssT)} \; e^{-\phi/2} = (2|n|g \,r/\kappa)
\sqrt{A_\ssZ/(2 A_\ssT)} $ and $m = m_\star \sqrt{A_\ssQ/A_\ssP}
\; e^{\phi/2} = [\kappa m_\star r/(2g)] \sqrt{A_\ssQ/A_\ssP}$, and
so $m \simeq 1/\rho_b \simeq 1/r$ and $\Delta U_\eff \simeq m^4$
if all other scales are equal.

However, just as for the case
we discussed in the previous sections, the most
interesting situation is where the brane field does not couple at
all to the bulk scalar: $a_t = a_z = a_p = a_q = 0$. In this case
if $A_\ssT \simeq A_\ssQ \simeq M^5$ and $A_\ssZ \simeq A_\ssP
\simeq M^3$ we have $\rho_b \simeq |n|/M$ and $m \lsim M$, and so
even though ${T_2}_0 \simeq M^4$ and $\Delta T_2 \simeq m^4$ are
as large as would generically be expected, both ${U_2}_0$ and
$\Delta U_\eff$ can vanish, regardless of how large $m$ and $M$
are.

From a geometrical point of view, this situation where the brane
tension is independent of the bulk dilaton is often obtained when
the brane is embedded in a supersymmetric bulk, since in this case
supersymmetry tends to require that the scalar be constant
everywhere in the bulk, and so naturally has a vanishing
derivative at the position of any source brane. The brane in
general breaks supersymmetry, but unless the bulk solution is
drastically modified (for example by bulk loops that may change
the classical extra-dimensional configuration), the couplings
$a_i$ between brane and bulk fields are expected to be small. The
previous discussion, then, ensures that $U_{\eff}$ is much smaller
than the physical mass of the particle integrated out.

We end our analysis with a discussion of how bulk loops can affect
the previous arguments.

\subsubsection*{Bulk loops}

But what about bulk loops? In particular, the above arguments
explicitly use the classical bulk equations when integrating out
the bulk KK modes to obtain $U_\eff$, and these can be expected to
be corrected by bulk loops.

When thinking about bulk loops it is useful to keep separate the
integration over KK modes whose wavelength is of order the size,
$\lambda \sim r$, of the extra dimensions and those of much
shorter wavelength. In particular, it is the long-wavelength modes
whose contributions can act over the size of the bulk and so
potentially modify in an important way the argument using the
classical bulk equations to derive $U_\eff$. We do not calculate
these here, but because these are the modes which dominantly
contribute to the Casimir energy in the extra dimensions we expect
from earlier explicit calculations \cite{CETori,CENontori} to find
that they generically contribute of order $1/r^4$ to the 4D vacuum
energy.

More dangerous are the contributions of the short-wavelength
modes, with $\lambda \ll r$, since these can potentially
contribute amounts of order $1/\lambda^4$ or $1/(\lambda^2 r^2)$
to $U_\eff$. However the effects of such short-wavelength modes
can be captured in terms of local terms in the low-energy bulk
effective action, and explicit calculations of the coefficients of
these terms \cite{UVSens} show that they are generically nonzero,
but cancel once summed over the field content of a 6D
supermultiplet.

But these explicit calculations do {\it not} apply to
short-wavelength bulk loops if the corresponding quantum
fluctuation occurs close to the branes, since in this case they
can instead contribute to local effective interactions in the
low-energy brane lagrangian \cite{CasBdy}, about whose general
form less is known. However, the order of magnitude of such
effects can be estimated using the calculations presented here, by
making an educated guess as to the size of their contribution to
the regularized action, $S_\reg$, as we now show.

Our main assumption when so doing is that each bulk loop comes
with a factor of $e^{2\phi}$, in addition to any factors of
$1/(2\pi)$ required by kinematics, if all indices are contracted
using the metric $\cg_{\ssM \ssN} = e^\phi g_{\ssM\ssN}$. This
loop counting follows from the fact, stated above, that this is
the frame for which $\phi$ enters undifferentiated into the
classical bulk supergravity action only as a pre-factor: $\L_\ssB
= e^{-2\phi} {\cal F}(\partial_\ssM \phi, \cg_{\ssM\ssN})$, and so
$e^{2\phi}$ plays the same role as does $\hbar$ in counting loops.
Notice that the scale-invariant choice for the regularized brane
action when written in this frame becomes
\be
 \L_\reg = - e^{-2\phi} \sqrt{-\cg} \; \left[ A_\ssT  +
 \frac12 \, A_\ssZ \,\cg^{mn} \partial_m \sigma \partial_n \sigma
  \right] \,,
\ee
showing that the tree level contribution for the brane action
arises with the same factor, $e^{-2\phi}$, as for the bulk action.

We therefore expect an $n$-loop contribution to $T_1$ and $Z_1$ in
this frame to be proportional to $e^{2(n-1)\phi}$, which leads to
the following loop expansion in the 6D Einstein frame:
\be
  T_1 \simeq e^{-\phi/2} \Bigl( A_\ssT^0 + A_\ssT^1 \, e^{2\phi} +
  \cdots \Bigr)
  \quad\hbox{and} \quad
 Z_1 \simeq e^{+\phi/2} \Bigl( A_\ssZ^0 + A_\ssZ^1 \, e^{2\phi} +
 \cdots \Bigr) \,.
\ee
Following the same steps as above then leads to the following
estimate for the leading corrections to the codimension-2 tension
coming from short-wavelength bulk loops:
\be
 \delta T_2(\phi) = C^0_\ssT + C^1_\ssT \, e^{2\phi} + \cdots \,,
\ee
where $C^0_\ssT$ and $C^1_\ssT$ are $\phi$-independent constants.
Consequently
\be
 \delta U_\eff \simeq -\frac12 \, \delta T_2'
 \simeq - C^1_\ssT \, e^{2\phi} \simeq - C^1_\ssT \left(
 \frac{\kappa}{ 2 g r} \right)^4 \,,
\ee
which again uses eq.~\pref{E:phircond} to trade $e^\phi$ for
$1/r$. Being of order $1/r^4$, is not systematically larger than
the contribution of longer-wavelength bulk loops.

Clearly the same estimates would argue that higher bulk loops are
also not dangerous, because all such loops are suppressed by even
more powers of the coupling $e^{\phi/2} \simeq \kappa/(2gr)$, that
can be {\it extremely} small when $r$ is large, such as is
required for the SLED proposal for approaching the cosmological
constant problem \cite{SLED,XDCC} (for which $2g/\kappa \simeq 10$
TeV while $1/r \simeq 10$ meV).

\section{Conclusions}

In this paper we analyzed effective theories for codimension two
branes, embedded in a higher dimensional space containing gravity
and a scalar field. In order to consistently define a coupling
between the brane and the bulk scalar, we represented the
codimension two source in terms of a regularizing codimension one
object, whose small size is determined by the dynamics of the
system. This procedure allowed us to define the tension of the
brane, called $T_2(\phi)$, and the low energy effective scalar
potential, indicated with $U_2(\phi)$, relevant below the
Kaluza-Klein scale.

We studied how the low energy scalar potential $U_2(\phi)$ is
sensitive to quantum corrections on the brane. In particular we
discussed under which conditions threshold effects, associated
with integrating out massive particles on the brane, are
suppressed in respect to naive expectations from dimensional
analysis. Threshold effects are reduced when the brane tension
$T_2(\phi)$ has little or no coupling to the bulk scalar. In this
case, although the brane tension $T_2(\phi)$ receives potentially
large corrections (of the order of the mass of the particle that
is integrated out), the size of the quantum corrected scalar
potential $U_2(\phi)$ that results is much smaller than
$T_2(\phi)$. This is in agreement with the known situation of
codimension two objects in pure gravity theories (for example
conical singularities), in which the brane tension is constant
(but non vanishing) and at the same time the low energy brane
potential is exactly zero allowing for flat on-brane geometries.
Our approach, in terms of a low energy effective theory, allows us
to go beyond the situation of pure gravity  and quantitatively
analyze how the coupling of the brane with bulk fields influences
the low energy potential.

As an illustration, we discussed how technical naturalness can be
achieved in a supersymmetric example, in which the extra
dimensional theory contains further degrees of freedom required by
supersymmetry. In our set-up we considered not only quantum
corrections to the low energy action due to brane threshold
effects. We also estimate quantum effects in the bulk, suggesting
that their contributions to the low energy effective potential can
be suppressed in respect to the brane ones.

The methods developed in this paper rely on the equations of
motion for bulk fields and on the brane junction conditions, and
offer a clear and intuitive geometrical interpretation of the
physics of how the bulk matches to codimension-2 branes, including
loop corrections. They allow a consistent derivation of effective
theories for higher codimension objects in a variety of cases, and
the analysis of their sensitivity to quantum effects on the brane
and in the bulk. We hope to further develop these topics in the
future, in particular in connection with supergravity models in
six dimensions.

\section*{Acknowledgements}
We wish to thank Claudia de Rham, Fernando Quevedo and Andrew
Tolley for many helpful comments and suggestions regarding
renormalization and naturalness with codimension-2 branes. CB's
research on this paper was partially supported by funds from the
Natural Sciences and Engineering Research Council (NSERC) of
Canada, CERN and McMaster University. Research at the Perimeter
Institute is supported in part by the Government of Canada through
NSERC and by the Province of Ontario through MRI.

\appendix

\section{The Explicit Quantum Calculation}
\label{app:EQC}

This appendix gives explicit details about the integrating out of
the heavy field $\psi$. Our starting point is eq.~\pref{Gammadef},
which defines the quantum action, $\Gamma(\phi,g)$, in terms of
the functional integral
\be \label{a:gammadef}
    \exp \Bigl[ i \Gamma(\phi,g) \Bigr] = \int {\cal D}\psi \;
    \exp \Bigl[ i S(\psi,\phi,g) \Bigr] \,.
\ee
This is most easily computed by first differentiating with respect
to $\phi$, giving
\ba
    \frac{\delta \Gamma}{\delta \phi} &=& e^{-i\Gamma} \int {\cal D} \psi \;
    \frac{\delta S}{\delta \phi} \, e^{iS} \nn\\
    &=& - \frac12 \sqrt{-g} \Bigl[ P'(\phi)
    \langle (\partial \psi)^2 \rangle + m_\star^2
    Q'(\phi) \, \langle \psi^2 \rangle
    \Bigr] \nn\\
    &=& - \frac12 \sqrt{-g} \left[ \frac{P'}{P} \,
    \langle P \, (\partial \psi)^2 \rangle + \frac{Q'}{Q}
    \, \langle m_\star^2 Q \, \psi^2 \rangle
    \right] \,,
\ea
where $\langle X(\psi) \rangle = e^{-i \Gamma} \int {\cal D} \psi
\, X(\psi) \, e^{iS}$. The problem reduces to computing $\langle P
(\partial \psi)^2 \rangle$ and $\langle m_\star^2 Q \, \psi^2
\rangle$, which can be obtained from the coincidence limit, $x'
\to x$, of the propagator, $G(x,x') = \langle \psi(x) \psi(x')
\rangle$. Because the result generically diverges in the
ultraviolet, we do so in $d$ spacetime dimensions and take $d =
4-2\epsilon$ at the end.

The calculation is most easily done with the canonically
normalized field, $\psi_\ssR = P^{1/2}(\phi) \psi$, and so
$P^{1/2} \partial_\mu \psi = \partial_\mu \psi_\ssR - \frac12
(P'/P) \partial_\mu \phi \, \psi_\ssR \equiv D_\mu \psi_\ssR$.
That is, specializing to constant $\phi$, write
\be
    \langle P(\phi) \psi(x) \psi(x') \rangle
    = \langle \psi_\ssR(x) \psi_\ssR(x') \rangle
    = -\frac{i}{L} \sum_{k=-\infty}^\infty \int
    \frac{\d^dp}{(2\pi)^d} \;
    \frac{e^{ip \cdot(x-x')}}{p^2 + q_k^2 +
    m^2} \,,
\ee
where $q_k = 2\pi k/L$ with $L = 2\pi \rho_b$ and $m^2 =
m^2(\phi)$ is given by eq.~\pref{mvsphi}. The coincidence limit of
this expression may be written
\be
    \langle P(\phi) \psi^2 \rangle =
    \langle \psi_\ssR^2 \rangle
    = \frac{1}{L} \int_0^\infty \d s
    \sum_{k =-\infty}^\infty \int \frac{\d^dp_\ssE}{(2\pi)^d} \;
    e^{-s(p^2 + q_k^2 + m^2)} \,,
\ee
which uses the identity $X^{-1} = \int_0^\infty \d s \; e^{-sX}$
and performs the Wick rotation to euclidean signature $\d^d p = i
\d^dp_\ssE$.

In this form the integrations over $p^\mu$ and the sum over $k$
may be performed explicitly, using the results
\be
    \int \d^dp_\ssE \; e^{-sp^2} =
    \left( \frac{\pi}{s} \right)^{d/2}
    \qquad \hbox{and} \qquad
    \sum_{k=-\infty}^\infty e^{-\pi k^2 t} = \vartheta_3(it) \,,
\ee
where $\vartheta_3(it)$ denotes the usual Jacobi theta-function.
Using these gives the expression
\ba \label{a:PsiSqExp}
    \langle P(\phi) \psi^2 \rangle =
    \langle \psi_\ssR^2 \rangle &=&
    \frac{1}{(4\pi)^{d/2} L} \int_0^\infty
    \frac{\d s}{s^{d/2}} \; e^{-s m^2} \,
    \vartheta_3 \left( \frac{4\pi i s}{L^2}
    \right) \nonumber\\
    &=& \frac{1}{4\pi L^{d-1}} \int_0^\infty
    \frac{\d t}{t^{d/2}} \; e^{- \lambda t}
    \, \vartheta_3 \left( it
    \right) \,,
\ea
where
\be
    \lambda(\phi) = \frac{m^2(\phi) L^2}{4\pi} =
    \frac{m_\star^2 L^2 Q(\phi)}{4\pi P(\phi)} \,.
\ee

We may repeat this calculation for the derivatives of $\psi$ to
compute $\langle P(\phi) \partial_m \psi \partial_n \psi \rangle$,
by evaluating $\partial_m \partial_n' G(x,x')$ and taking the
limit $x' \to x$. This amounts to inserting a factor of $ip_m (-i
p_n) = p_m p_n$ into the integrand of the appropriate expression.
The integral over $p$ and the sum over $k$ may again be performed,
using
\ba
    \int \d^dp_\ssE \; p_\mu p_\nu \, e^{-s p^2} &=& \eta_{\mu\nu}
    \left( \frac{\pi}{s} \right)^{d/2} \frac{1}{2s}
    \nonumber\\
    \sum_{k=-\infty}^\infty k^2 e^{-\pi k^2 t} &=& -\frac{1}{\pi}
    \frac{\partial}{\partial t} \sum_{k=-\infty}^\infty
    e^{-\pi k^2 t} = -\frac{i}{\pi} \vartheta_3'(it) \,.
\ea
Here the prime on $\vartheta_3$ denotes differentiation with
respect to its argument $\tau = it$. With these expressions we
have
\be \label{a:dPsiSqExp}
    \langle P(\phi) \partial_\mu \psi \partial_\nu \psi \rangle
    = \langle \partial_\mu \psi_\ssR \partial_\nu \psi_\ssR \rangle
    = \eta_{\mu\nu} \, \frac{1}{2L^{d+1}}
    \int_0^\infty \frac{\d t}{t^{(d+2)/2}} \; e^{-\lambda t}
    \vartheta_3(it) \,,
\ee
and
\be \label{a:dzPsiSqExp}
    \langle P(\phi) (\partial_z \psi)^2 \rangle =
    \langle (\partial_z \psi_\ssR)^2 \rangle =
    - \frac{i}{L^{d+1}} \int_0^\infty \frac{\d t}{t^{d/2}}
    \; e^{-\lambda t} \vartheta_3'(it) \,.
\ee

The problem is reduced to the evaluation of the following two
one-dimensional integrals:
\be
    I_\alpha(\lambda) := \int_0^\infty \frac{\exd t}{t^\alpha}
    \, e^{-\lambda t} \, \vartheta_3(it) \,,
\ee
and
\be
    J_\alpha(\lambda) := i\int_0^\infty \frac{\exd t}{t^\alpha}
    \, e^{-\lambda t} \, \vartheta_3'(it)
    \,,
\ee
whose properties are explored in some detail in Appendix
\ref{app:functions}.

\subsection*{An important identity}

The properties of the integrals $I_\alpha(\lambda)$ and
$J_\alpha(\lambda)$ imply an important identity,
\be \label{a:ExpIdentityapp}
    \langle {\cal X} \rangle := \Bigl\langle P(\phi) \,
    \partial_\ssM \psi \partial^\ssM \psi
    + m_\star^2 Q(\phi) \, \psi^2 \Bigr\rangle = 0 \,,
\ee
which holds in dimensional regularization. This result follows
from eqs.~\pref{a:PsiSqExp}, \pref{a:dPsiSqExp} and
\pref{a:dzPsiSqExp}, written in the form
\ba \label{a:expvalresults}
    \langle m_\star^2 Q(\phi) \, \psi^2 \rangle &=&
    \frac{\lambda}{L^{d+1}} \, I_{d/2}(\lambda) \nn\\
    \langle P(\phi) \, \partial_\mu\psi \partial^\mu
    \psi \rangle &=&
    \frac{d}{2\, L^{d+1}} \, I_{1+d/2}(\lambda) \nn\\
    \langle P(\phi) \, (\partial_z\psi)^2 \rangle &=&
    - \frac{1}{L^{d+1}} \, J_{d/2}(\lambda) \,,
\ea
which allow eq.~\pref{a:ExpIdentityapp} to be written
\be
    \langle {\cal X} \rangle = \frac{1}{L^{d+1}} \Bigl[
    \lambda I_{d/2}(\lambda) +
    \frac d2 \, I_{1+d/2}(\lambda)
    - J_{d/2}(\lambda) \Bigr] \,.
\ee
This combination is shown to vanish identically in Appendix
\ref{app:functions}.

\subsection*{The massless limit}

Finally, we evaluate the stress energy, $\Bigl\langle T_{mn}
\Bigr\rangle = \Bigl \langle P(\phi) \, \partial_m \psi \,
\partial_n \psi \Bigr \rangle$, explicitly in the limit when $m_\star =
0$. In this limit the stress energy is given by
\be
    \Bigl\langle T_{\mu\nu} \Bigr\rangle
     = \frac{C_d}{L^{d+1}} \; \eta_{\mu\nu} \quad
     \hbox{and} \quad
    \Bigl\langle T_{zz} \Bigr\rangle
     = \frac{C_z}{L^{d+1}}
    \,,
\ee
where $C_d = \frac12 \, I_{1+d/2}(0)$ and $C_z = - J_{d/2}(0) = -
\frac d2 \, I_{1+d/2}(0)$. Keeping in mind that
$I_\alpha^\infty(0) = 0$ when $\hbox{Re}\, \alpha > \frac12$ (and
so in particular when $\alpha = 1+d/2$), this gives (using the
results of Appendix \ref{app:functions})
\be
    -\frac{C_z}{d} = C_d = \frac12 \; I_{1+d/2}^f(0)
    = \frac{1}{\pi^{(d+1)/2}} \; \Gamma\left( \frac{d+1}{2}
    \right) \zeta(d+1)
    \,,
\ee
which may be explicitly evaluated when $d=4$ to give $C_4 = 3\,
\zeta(5) / (4\pi^2) \simeq 7.68$.

\section{The functions $I_\alpha(z)$ and $J_\alpha(z)$}
\label{app:functions}

The previous appendix shows the utility of defining the following
functions
\be
    I_\alpha(\lambda) := \int_0^\infty \frac{\exd t}{t^\alpha}
    \, e^{-\lambda t} \, \vartheta_3(it) \,,
\ee
and
\be
    J_\alpha(\lambda) := i\int_0^\infty \frac{\exd t}{t^\alpha}
    \, e^{-\lambda t} \, \vartheta_3'(it)
    \,,
\ee
where, as before, the prime denotes differentiation with respect
to $\tau = it$. This appendix collects many useful properties of
these two functions.

\subsection*{Ultraviolet divergent parts}

To understand the convergence of the integrals we require the
following asymptotic forms \cite{WW} for $\vartheta_3(it)$,
\ba
    &&\vartheta_3(it) = 1 + 2\, e^{-\pi t} + \cdots \qquad
    \qquad \hbox{when $t \to \infty$} \nn\\
    \hbox{and} \qquad
    &&\vartheta_3(it) = \frac{1}{\sqrt{t}} \Bigl[ 1 +
    2\, e^{-\pi/t} +
    \cdots \Bigr] \quad\;\; \hbox{when $t \to 0$} \,.
\ea
These imply that the integral defining $I_\alpha(\lambda)$
converges as $t \to \infty$ for any $\alpha$ when Re$\,\lambda >
0$, and for Re$\,\alpha > 1$ if Re$\,\lambda = 0$. By contrast,
the exponential falloff of the function $\vartheta_3'(it)$ for
large $t$ ensures the integral defining $J_\alpha(\lambda)$
converges for large $t$ for any $\alpha$ if Re$\,\lambda > -\pi$.
On the other hand, convergence of $I_\alpha(\lambda)$ for $t \to
0$ requires Re$\,\alpha < \frac12$, while the small-$t$
convergence of $J_\alpha(\lambda)$ requires Re$\,\alpha < -
\frac12$.

Our eventual applications make us particularly interested in the
cases where $\alpha = d/2$ or $\alpha = (d + 2)/2$, where $d$ is a
positive integer (with $d=4$ being particularly interesting).
Although Re$\,\lambda > 0$ is sufficient to ensure the convergence
of the integrals for all $\alpha$ as $t \to \infty$, the above
asymptotic forms show that $I_\alpha(\lambda)$ in general diverges
as $t \to 0$ for all $d$ of interest. This divergence represents
the ultraviolet divergence of the physical quantities under study.

It is useful to isolate this divergence by writing $I_\alpha =
I^\infty_\alpha + I^f_\alpha$ (and ditto for $J_\alpha$), where
the `infinite' parts are obtained by replacing $\vartheta_3(it)
\to 1/\sqrt{t}$ and the `finite' parts are obtained by replacing
$\vartheta_3(it) \to \vartheta_3(it) - 1/\sqrt{t}$. Regularizing
the divergent parts using dimensional regularization then gives
\be
    I^\infty_\alpha(\lambda) = \int_0^\infty \frac{\d
    t}{t^{\alpha + \frac12}} \; e^{-\lambda t}
    = \lambda^{\alpha - \frac12} \,
    \Gamma\left( \frac12 - \alpha \right) \,,
\ee
and
\be
 J^\infty_\alpha(\lambda) = -\frac12 \int_0^\infty
 \frac{\d t}{t^{\alpha + \frac32}} \; e^{- \lambda t}
 = - \frac12 \, \lambda^{\alpha +
    \frac12} \, \Gamma \left( - \frac12 - \alpha \right) \,,
\ee
where $\Gamma(z)$ is Euler's generalized factorial function,
satisfying $z\Gamma(z) = \Gamma(z+1)$ and $\Gamma(k+1) = k!$ for
$k$ a nonnegative integer. Notice that if $\alpha = {d}/{2}$ or
$\alpha = (d + 2)/2$, this expression has poles when $d$ is
positive and odd (and so when the total dimension of spacetime on
the codimension-1 brane, $d+1$, is even). As is often the case in
dimensional regularization, the one-loop divergences happen to be
finite when $d$ is even and positive (so $d+1$ is odd), and in
particular for the cases of practical interest: $\alpha = {d}/{2}
= 2 - \epsilon$ or $\alpha = ({d}/{2}) + 1 = 3 - \epsilon$.

Once these are taken out the remaining integrals
\ba
 I^f_\alpha(\lambda) &:=& \int_0^\infty \frac{\exd t}{t^\alpha}
    \, e^{-\lambda t} \, \left[ \vartheta_3(it) -
    \frac{1}{\sqrt{t}} \right] \\
 J^f_\alpha(\lambda) &:=& \int_0^\infty \frac{\exd t}{t^\alpha}
    \, e^{-\lambda t} \, \left[ i\vartheta_3'(it) +
    \frac{1}{2 \, t^{3/2}} \right]
    \,, \nn
\ea
converge exponentially for small $t$.

\begin{figure}[ht]
\begin{center}
\epsfig{file=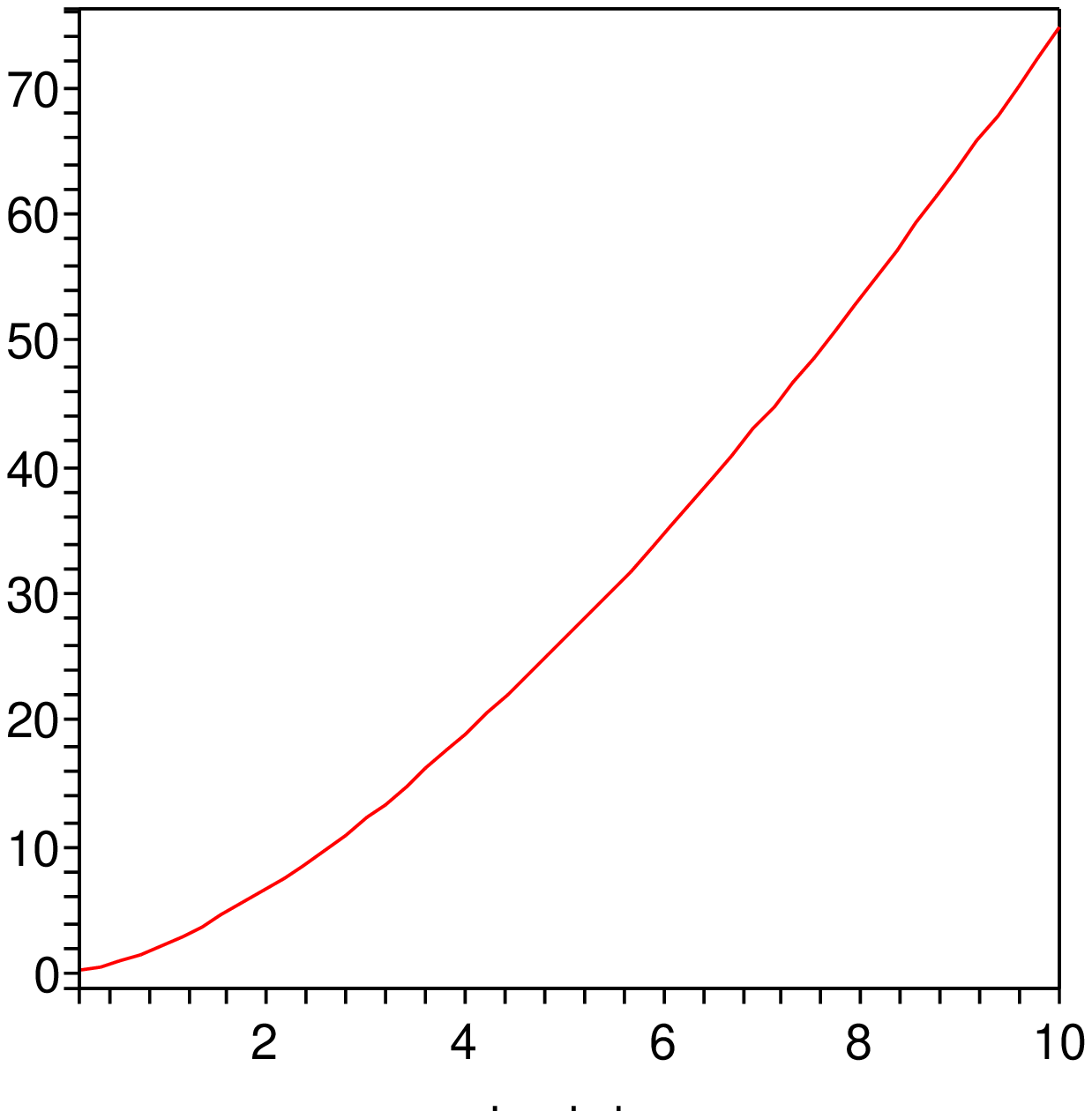, height=70mm,width=67mm}
\epsfig{file=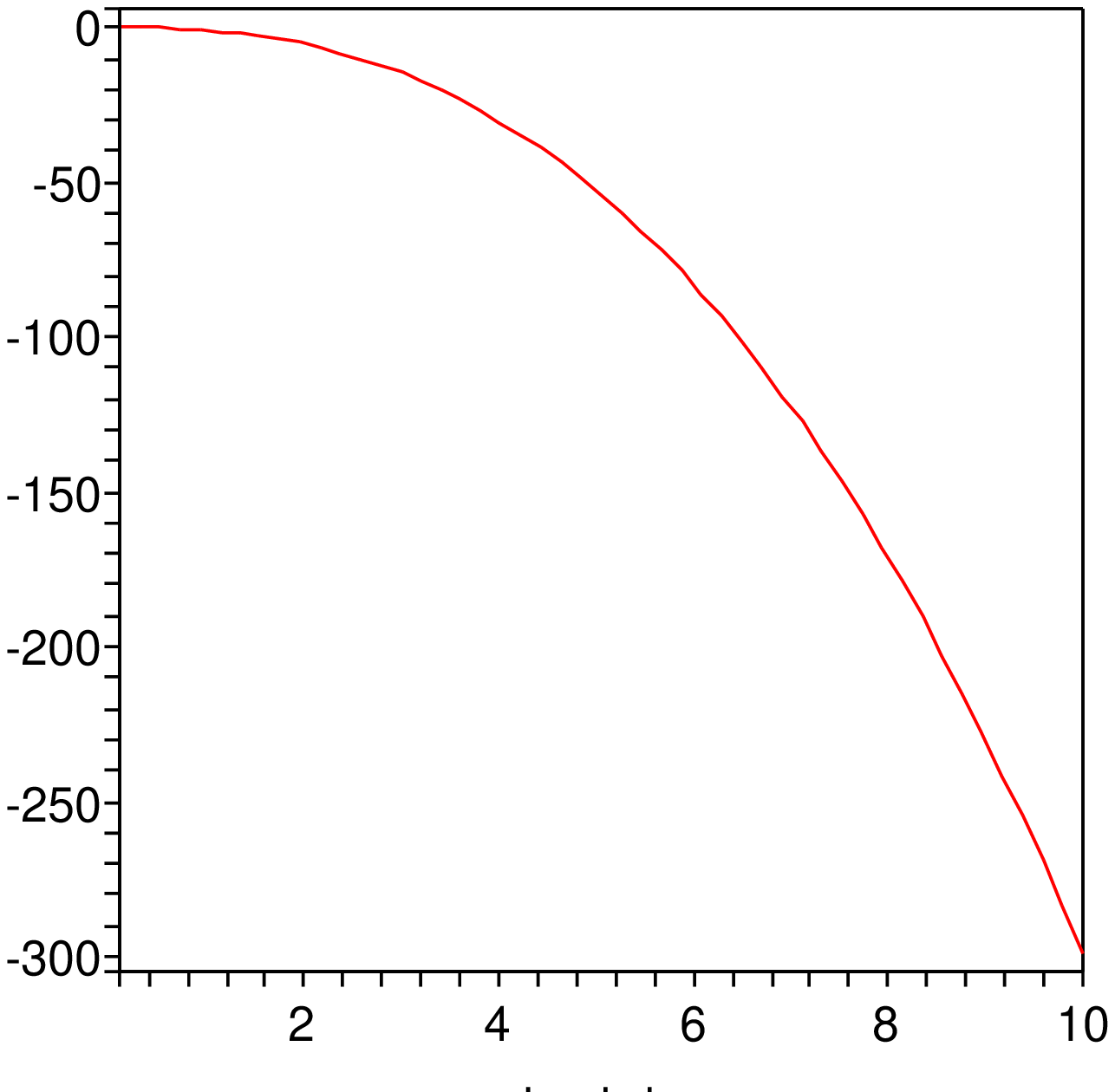, height=70mm,width=67mm} \caption{A plot of
the functions $I_2(\lambda)$ (left) and $I_3(\lambda)$ (right) vs
$\lambda$.}
   \label{fig:I23plot}
\end{center}
\end{figure}

\subsection*{A useful identity}

A very useful property of these integrals follows by integrating
by parts in the definition of $J_\alpha(\lambda)$, leading to
\be \label{a:intbyparts}
    J_\alpha(\lambda) = \lambda I_\alpha(\lambda) + \alpha I_{\alpha + 1}(\lambda)
    \qquad \hbox{if} \quad \hbox{Re}\, \alpha < -\frac12 \,,
    \quad \hbox{Re} \, \lambda > 0
    \,.
\ee
Here the assumptions for Re$\,\alpha$ and Re$\,\lambda$ are
required to ensure the vanishing of the surface term. Since this
identity proves very useful in the main text, we now show that it
applies even for $\alpha$ not in the above regions, provided that
the divergences encountered are dimensionally regularized.
Although we now demonstrate this in detail, the conclusion also
follows from eq.~\pref{a:intbyparts} by analytic continuation in a
potentially wider set of regularization schemes.

We wish to show that the following quantity vanishes:
\be
    \langle {\cal X} \rangle = \frac{1}{L^{d+1}} \Bigl[
    \lambda I_{d/2}(\lambda) +
    \frac d2 \, I_{1+d/2}(\lambda)
    - J_{d/2}(\lambda) \Bigr] \,.
\ee
To this end notice first that the potentially divergent parts
cancel identically from this combination, since
\be
    \lambda I^\infty_{d/2} +
    \frac d2 \, I^\infty_{1+d/2}
    - J^\infty_{d/2} =
    \lambda^{d+1/2} \left[
    \left( \frac d2 + \frac12 \right) \Gamma\left(
    -\frac12 - \frac d2 \right) + \Gamma \left(
    \frac12 - \frac d2 \right) \right] \,,
\ee
which vanishes by virtue of the identity $z \Gamma(z) =
\Gamma(z+1)$, specialized to $z = -\frac12 - \frac d2$. The finite
parts similarly cancel, since they may be written
\ba
    \lambda I^f_{d/2} +
    \frac d2 \, I^f_{1+d/2}
    - J^f_{d/2} &=&
    \int_0^\infty \frac{\exd t}{t^{1+d/2}} \; e^{-\lambda t}
    \left[ \left( \lambda t + \frac d2 \right)
    \left( \vartheta_3 - \frac{1}{\sqrt{t}} \right)
    - t \left( \frac{\exd \vartheta_3}{\exd t} +
    \frac{1}{2\, t^{3/2}} \right) \right] \nn\\
    &=& - \int_0^\infty \exd t \; \frac{\exd}{\exd t}
    \left[ \frac{e^{-\lambda t}}{t^{d/2}} \left(
    \vartheta_3 - \frac{1}{\sqrt{t}} \right) \right]  \,,
\ea
which vanishes because the integrand vanishes exponentially
quickly as both $t \to 0$ and $t \to \infty$. (If $\lambda = 0$
then the limit $t \to \infty$ still vanishes like a power of $1/t$
provided $d > 0$.)

\subsection*{The special case $\lambda = 0$}

The case where $I_\alpha^f$ is evaluated at $\lambda = 0$ arises
in the main text, and can be evaluated explicitly. In this case we
have
\ba
    I_\alpha^f(0) &=& \int_0^\infty \frac{\exd t}{t^\alpha} \,
    \left[ \vartheta_3(it) - \frac{1}{\sqrt{t}} \right] \nn\\
    &=& \int_0^\infty \exd t \; t^{\alpha - \frac32}
    \Bigl[ \vartheta_3(it) - 1 \Bigr] \,,
\ea
where we change variables $t \to 1/t$ and use the identity
\be
    \vartheta_3\left( \frac{i}{t} \right) = \sqrt{t} \;
    \vartheta_3(it) \,.
\ee
The remaining integral may be evaluated using the identity,
\be
    \int_0^\infty \frac{\exd t}{t} \; t^{s/2} \Bigl[
    \vartheta_3(it) - 1 \Bigr] = \frac{2}{\pi^{s/2}}
    \; \Gamma \left( \frac{s}{2} \right)
    \zeta(s) \,,
\ee
where $\zeta(s)$ is Riemann's zeta function, to give
\be
    I_\alpha^f(0) = \frac{2}{\pi^{\alpha - \frac12}} \;
    \Gamma\left( \alpha - \frac12 \right)
    \zeta \Bigl( 2 \alpha -1
    \Bigr) \,.
\ee

\subsection*{The special case $\alpha = 2$}

The full expression for $I_\alpha^f(\lambda)$ may also be obtained
for the special value $\alpha = 2$, which is of practical interest
in the case $d =4$. In this case we use the definition
\be
    \vartheta_3(it) = \sum_{k=-\infty}^\infty e^{-\pi k^2 t}
\ee
and the great convergence properties of the sums and integrals to
reverse the order of summation and integration, leading to
\ba
    I_\alpha^f(\lambda) &=& \int_0^\infty \frac{\exd t}{t^\alpha}
    \left[ \vartheta_3(it) - \frac{1}{\sqrt{t}} \right] \,
    e^{-\lambda t} \nn\\
    &=& \int_0^\infty \exd t \; t^{\alpha - \frac32}
    \Bigl[ \vartheta_3(it) - 1 \Bigr] \,
    e^{-\lambda/ t} \nn\\
    &=& 2 \sum_{n=1}^\infty \int_0^\infty \exd t \; t^{\alpha - \frac32}
    \; e^{ - \pi n^2 t -\lambda/ t } \nn\\
    &=& 4 \pi^{(1- 2\alpha)/4} \lambda^{(2\alpha - 1)/4}
    \sum_{n=1}^\infty n^{(1-2\alpha)/2} K_{\alpha-\frac12}\Bigl(
    \sqrt{4\pi n^2 \lambda} \Bigr) \,,
\ea
where $K_\nu(z)$ denotes the modified Bessel function of the
second kind, $K_\nu(z) = Y_{i\nu}(z)$. In the case $\alpha =2$
this sum can be performed explicitly in terms of the Digamma
function,
\be
    \hbox{\rm Li}_s(z) \equiv \sum_{n=1}^\infty \frac{z^n}{n^s}
    \,,
\ee
to give
\be
    I_2^f(\lambda) = \frac{1}{\pi} \left[ \sqrt{4\pi \lambda} \;
    \hbox{\rm Li}_2\left( e^{-\sqrt{4\pi\lambda}} \right) +
    \hbox{\rm Li}_3\left(e^{-\sqrt{4\pi\lambda}} \right) \right]
    \,.
\ee

\subsection*{Asymptotic forms}

To identify asymptotic forms for large and small $\lambda$ we
write $F(t) \equiv \vartheta_3(it) - t^{-1/2}$, in terms of which
the finite integral becomes
\ba
    I^f_\alpha(\lambda) &=& \lambda^{\alpha -1}
    \int_0^\infty \frac{\exd u}{u^\alpha} \,
    e^{-u} F(u/\lambda) \nn\\
    &\approx& 2 \,\lambda^{\alpha -\frac12}
    \int_0^\infty \exd u \; u^{-\alpha-\frac12} \,
    e^{-u - \pi\lambda/u} \propto \lambda^{\alpha -\frac12}
    (\pi \lambda)^{-\alpha -
    \frac14} e^{-2\sqrt{\pi\lambda}} \qquad
    \hbox{when $\lambda \gg 1$} \nn\\
    &\approx& \lambda^{\alpha -1}
    \int_0^\infty \exd u \; u^{-\alpha} \,
    e^{-u} = \lambda^{\alpha -
    1} \Gamma\Bigl[1-\alpha\Bigr] \qquad
    \hbox{when $\lambda \ll 1$} \,.
\ea
This uses the asymptotic forms $F(u/\lambda) \approx 2 \,
e^{-\pi\lambda/u} \sqrt{\lambda/u}$ when $\lambda \gg 1$ and
$F(u/\lambda) \approx 1$ when $\lambda \ll 1$. The large-$\lambda$
limit is evaluated using the saddle-point approximation, for which
\be
    \int \exd u \, f(u) \, e^{-h(u)} \propto
    \frac{f(u_c)}{\sqrt{ h''(u_c)}} \,
    e^{-h(u_c)} \,,
\ee
where $u_c$ is defined by the condition $h'(u_c) = 0$. This is
$u_c = \sqrt{\pi\lambda}$ in the case of interest, for which $h(u)
= u + \pi\lambda/u$.

The other integral of interest is
\be
    J_\alpha^f(\lambda) \equiv
    \int_0^\infty \frac{\exd t}{t^\alpha}
    \, e^{-\lambda t} \, \left[ i\vartheta_3'(it)
    + \frac{1}{2 \, t^{3/2}} \right] \,,
\ee
so writing $G(t) \equiv i\vartheta_3'(it) + (2 \, t^{3/2})^{-1}$
the integral becomes
\ba
    J^f_\alpha(\lambda) &=& \lambda^{\alpha -1}
    \int_0^\infty \frac{\exd u}{u^\alpha} \,
    e^{-u} G(u/\lambda) \nn\\
    &\approx& - \lambda^{\alpha + \frac12}
    \int_0^\infty \exd u \; u^{-\alpha-\frac32} \,
    e^{-u - \pi\lambda/u} \propto \lambda^{\alpha + \frac12}
    (\pi \lambda)^{-\alpha -
    \frac54} e^{-2\sqrt{\pi\lambda}} \qquad
    \hbox{when $\lambda \gg 1$} \nn\\
    &\approx& \frac12 \, \lambda^{\alpha + \frac12}
    \int_0^\infty \exd u \; u^{-\alpha - \frac32} \,
    e^{-u} = \frac12 \, \lambda^{\alpha +
    \frac12} \Gamma\left[-\frac12 - \alpha \right] \qquad
    \hbox{when $\lambda \ll 1$} \,,
\ea
which uses the asymptotic forms $G(u/\lambda) \approx -
(\lambda/u)^{3/2} e^{-\pi\lambda/u}$ when $\lambda \gg 1$ and
$G(u/\lambda) \approx \frac12 (\lambda/u)^{3/2}$ when $\lambda \ll
1$.

\subsubsection*{Infrared singularities for small $\lambda$}

The small-$\lambda$ limit involves some subtleties when $\alpha$
is in the regime of practical interest, $\alpha = \frac d2 =
2-\epsilon$. Naively specializing the above asymptotic limits to
this case gives
\be
 I_2^f(\lambda) \sim \lambda \Gamma(-1+\epsilon) =
 -\frac{\lambda}{\epsilon} + O(1) \,,
\ee
which diverges as $\epsilon \to 0$. Because we know that $I^f_2$
converges absolutely for nonzero positive $\lambda$ by
construction, this divergence in the small-$\lambda$ limit
represents an infrared mass singularity for small $m$ which
invalidates an expansion in powers of $\lambda$.

To isolate this singularity explicitly, it is worth multiply
differentiating the integral expression for $I_2^f(\lambda)$ with
respect to $\lambda$, to obtain
\ba
    \frac{\exd^2 I^f_2}{\exd \lambda^2} &=&
    \int_0^\infty {\exd t} \, e^{-\lambda t}
    \left[ \vartheta_3(it) - \frac{1}{\sqrt{t}} \right]
    = \frac{1}{\lambda} \int_0^\infty \exd u \, e^{-u} \,
    F(u/\lambda) \nn\\
    &\approx& \frac{1}{\lambda} \int_0^\infty \exd u
    \, e^{-u} = \frac{1}{\lambda} \quad \hbox{when $\lambda
    \ll 1$}\,,
\ea
which when integrated implies $I^f_2(\lambda) \approx \lambda(\ln
\lambda - 1) + A\lambda + B$ when $\lambda \ll 1$, where $A$ and
$B$ are integration constants.

The constants $A$ and $B$ may be obtained by going back to the
original integral defining $I^f_2(\lambda)$ and numerically
integrating in the small-$\lambda$ limit. This leads to
\be
    A = \left[ \frac{\exd I^f_2}{\exd \lambda}
    - \ln \lambda \right]_{\lambda = 0}
    = -\int_0^\infty \frac{\exd t}{t}
    \, \left[ \vartheta_3(it) - \frac{1}{\sqrt{t}}
    - \frac{t}{1+t} \right]
    \simeq -1.94 \,,
\ee
which uses the representation
\be
    \ln \lambda = \int_0^\infty \exd u \left[\frac{1}{u+1} -
    \frac{1}{u+\lambda} \right] = (\lambda - 1) \int_0^\infty
    \frac{\exd t }{(1 + t) (1 + \lambda t)} \,,
\ee
where $t = 1/u$. Similarly,
\be
    B = I^f_2(0) = \int_0^\infty \frac{\exd t}{t^2}
    \, \left[ \vartheta_3(it) - \frac{1}{\sqrt{t}} \right]
    = \frac{\zeta(3)}{\pi} \simeq 0.38 \,.
\ee

\section{Supergravity Equations of Motion}
\label{app:sugra}

This appendix summarizes the equations of motion for the bosonic
part of 6D chiral gauged supergravity, and uses these to trace how
the arguments of the main text change when applied to this case.
The action for the theory is given (in the 6D Einstein frame and
for the case of vanishing hyperscalars --- $\Phi^i = 0$) by
\cite{NS}
\ba \label{6DSugraAction1}
    \frac{{\cal L}_\ssB}{\sqrt{-g}} &=& - \frac{1}{2 \kappa^2} \,
    g^{\ssM\ssN} \Bigl[ R_{\ssM\ssN} + \partial_\ssM \phi \,
    \partial_\ssN \phi \Bigr] - \frac{2g^2}{\kappa^4} \; e^\phi \nn\\
    && \qquad\qquad - \frac14 \, e^{-\phi} \, F_{\ssM\ssN} F^{\ssM\ssN}
    - \frac{1}{2 \cdot 3!} \, e^{-2\phi} \, G_{\ssM\ssN\ssP}
    G^{\ssM\ssN\ssP} \,,
\ea
where we specialize to a single gauge field, $F_{\ssM\ssN} =
\partial_\ssM A_\ssN - \partial_\ssN A_\ssM$ and Kalb-Ramond field,
$G_{\ssM\ssN\ssP} = \partial_\ssM B_{\ssN\ssP} +
\partial_\ssN B_{\ssP\ssM} + \partial_\ssP B_{\ssM\ssN} + (A_\ssP
F_{\ssM\ssN} \, \hbox{terms})$. $g$ and $\kappa$ are coupling
constants that respectively have dimension (mass)${}^{-1}$ and
(mass)${}^{-2}$.

The field equations obtained from this action are:
\ba \label{6DSugraEqns}
    \Box \, \phi + \frac{\kappa^2}{6} \, e^{-2 \phi} \;
    G_{\ssM\ssN\ssP} G^{\ssM\ssN\ssP} +
    \frac{\kappa^2}{4} \, e^{-\phi} \; F_{\ssM\ssN} F^{\ssM\ssN}
    - \frac{2 \,g^{2}}{\kappa^2} \, e^{\phi} &=& 0
    \qquad \hbox{(dilaton)}\nn \\
    D_\ssM \Bigl(e^{ -2 \phi} \, G^{\ssM\ssN\ssP} \Bigr) &=& 0 \qquad
    \hbox{(2-Form)} \nn\\
    D_\ssM \Bigl(e^{ - \phi} \, F^{\ssM\ssN} \Bigr) + e^{-2 \phi}
    \, G^{\ssM\ssN\ssP} F_{\ssM\ssP} &=& 0 \qquad \hbox{(Maxwell)}\nn\\
    R_{\ssM\ssN} + \partial_\ssM\phi \, \partial_\ssN\phi
    + \frac{\kappa^2}{2} \, e^{-2 \phi}
    \; G_{\ssM\ssP\ssQ} {G_\ssN}^{\ssP\ssQ} + \kappa^2 e^{- \phi} \;
    F_{\ssM\ssP} {F_\ssN}^\ssP
    + \frac12 \, (\Box\,
    \phi)\, g_{\ssM\ssN} &=& 0 \,.\quad\;\; \hbox{(Einstein)}\nn\\
\ea
An important feature of these equations is their invariance under
the replacement \cite{Witten}
\be
    e^\phi \to \lambda e^\phi \qquad \hbox{and} \qquad
    g_{\ssM\ssN} \to \lambda^{-1} g_{\ssM\ssN} \,,
\ee
with all other fields held fixed. Also notice that evaluating the
action, eq.~\pref{6DSugraAction1} using the dilaton and Einstein
equations of eqs.~\pref{6DSugraEqns}, implies the action evaluates
to
\be \label{SBSGclass}
 S_{\ssB\,{cl}} =  S_\GH +
 \frac{1}{2\kappa^2} \int \exd^6x \sqrt{- g_{cl}}
 \;  \Box \, \phi_{cl} \,,
\ee
where $S_\GH$ denotes the Gibbons-Hawking term, as in the main
text.

\subsection*{Compactified solutions}

For static solutions compactified to two dimensions supported by
Maxwell flux our interest is in field configurations of the form
\be \label{diagmetricform}
    \exd s^2 =  e^{2 W} \,
    \eta_{\mu\nu} \, \exd x^\mu \exd x^\nu
    + \exd \rho^2 + e^{2 B}
    \exd \theta^2 \quad \hbox{and} \quad
    A_\ssM \, \exd x^\ssM = A \, \exd \theta \,,
\ee
with component functions, $W$, $B$, $\phi$ and $A$, depending only
on $\rho$. Denoting differentiation with respect to $\rho$ by
primes, the field equations reduce to the following set of coupled
partial differential equations. The Maxwell equation is:
\be
    A'' + ( 4W' -B' -\phi') A' =
    e^{B-4W + \phi} \left( e^{-B + 4W - \phi} A' \right)' = 0 \,,
\ee
The dilaton equation is:
\be
    \phi'' + (4W' + B' ) \phi' +
    \frac{\kappa^2}{2} \, e^{- 2B -\phi} (A')^2
    - \frac{2 g^2}{\kappa^2} \, e^{\phi} =0 \,,
\ee
The $(\mu\nu)$ Einstein equation is:
\be
    W'' + W'(4 \, W' + B') - \frac{\kappa^2}{4} \,
    e^{- 2B -\phi} (A')^2
    + \frac{g^2}{\kappa^2}\, e^{\phi}
    = 0 \,.
\ee
The $(\rho\rho)$ Einstein equation is:
\be
    4\, W'' + B'' + 4 \, (W')^2 + (B')^2 + (\phi')^2
    + \frac{3\kappa^2}{4} \,
    e^{- 2B - \phi} (A')^2
    + \frac{g^2}{\kappa^2} e^{\phi} = 0 \,.
\ee
The $(\theta\theta)$ Einstein equation is:
\be
    B'' + B' (4\, W' + B' ) + \frac{3\kappa^2}{4} \,
    e^{- 2B - \phi} (A')^2
    + \frac{g^2}{\kappa^2}\, e^{\phi} = 0 \,.
\ee

Notice that the combination $4(\mu\nu)+(\theta\theta) -
(\rho\rho)$ of the three Einstein equations can be rewritten as
the constraint,
\be \label{newconstraint}
 12 (W')^2 + 8 W' B' - (\phi')^2 - \kappa^2 \left( e^{-4W -
 \phi/2} f \right)^2 + \frac{4 \, g^2}{\kappa^2} \, e^\phi = 0 \,,
\ee
which uses the solution to the Maxwell equation, $A' = f \,
e^{B-4W+\phi}$, with $f$ constant. This differs from the
constraint obtained for the pure massless scalar-tensor theory of
the main text only by the last two terms.

\subsection*{Jump conditions}

Using the same choice for the regularized brane action as in the
main text, eq.~\pref{Sregdef}, implies the same junction
conditions as were found there, eqs.~\pref{cod2match}:
\ba \label{app:cod2match}
 &&\lim_{\rho \to 0} \left( e^{B + d W} \partial_\rho \phi
 \right) = \frac{\kappa^2 T_2'}{2\pi} \,, \quad
 \lim_{\rho \to 0} \left( e^{B + d W} \partial_\rho W
 \right) = \frac{\kappa^2 U_2}{2\pi d} \\
 \hbox{and} \quad
 &&\lim_{\rho \to 0} \left( e^{B + d W} \partial_\rho B
 \right) = 1 - \frac{\kappa^2}{2\pi} \left[ T_2 + \left(
 \frac{d-1}{d} \right) U_2 \right] \nn\,.
\ea
with
\be
    T_2 = - \left( \frac{2 \pi \rho_b}{4} \right)
    \hg^{\mu\nu} T_{\mu\nu}
    \quad \hbox{and} \quad
    U_2 = 2 \pi \rho_b \, \hg^{\theta\theta}
    T_{\theta\theta} \,,
\ee
as before (using $d=4$).

The important new difference is that the quantity $U_2$ that
appears here is {\it not} related to the brane contribution to the
very-low-energy effective potential by $U_\eff = \frac{1}{d}
\sum_b U_2$, because the bulk action satisfies
eq.~\pref{SBSGclass} instead of $S_\ssB = S_\GH$. As a
consequence, classically integrating out the bulk KK modes in this
case instead gives (with $d=4$) \cite{BdRHT}
\ba \label{app:newUeff}
 \L_\eff(\phi,\rho_b) &=& 2\pi \rho_b
 \sum_b \left[ \L_\reg
 - \frac12 \, \hg_{mn} \; \frac{\partial \L_\reg}{
 \partial \hg_{mn}}
 - \frac12 \, \frac{\partial \L_\reg}{\partial \phi}
 \right] \nn\\
 \hbox{or} \quad
 U_\eff &=& \sum_b \left( \frac{U_2}{4}
 - \frac{T_2'}{2} \right) \,,
\ea
and precisely the same for $\Delta \L_\eff$ as a function of
$\Delta \L_\reg$ and its derivatives.

\subsection*{Constraint}

The constraint relating $U_2$ and $T_2$ is now derived by
eliminating the derivatives $W'$, $B'$ and $\phi'$ using the jump
conditions. $A'$ can similarly be related to the corresponding
brane current, $\delta S_\reg/\delta A_\ssM$, using its jump
condition \cite{UVCaps,BdRHT}. However, since the last two terms
of eq.~\pref{newconstraint} are suppressed relative to the first
three by positive powers of $\rho_b$ they may be neglected for
small $\rho_b$, as can contributions of order $\rho_b^2 R$
\cite{BdRHT}. As a consequence $U_2$ and $T_2$ are related to one
another by the same constraint, eq.~\pref{constraint}, as was
derived in the main text for massless scalar-tensor gravity:
\be \label{app:constraint}
 U_2 \left[ \frac{4\pi}{\kappa^2} - 2 T_2 - \left(
 \frac{d-1}{d} \right) U_2 \right] - \left( T_2' \right)^2
 \simeq 0 \,.
\ee
This implies that $\rho_b(\phi)$ is to good approximation obtained
by the same condition, $U_2 \simeq 0$, after which
eq.~\pref{app:newUeff} gives $U_\eff(\phi)$ with $U_2(\phi) =
U_2(\phi,\rho_b(\phi))$ and $T_2(\phi) = T_2(\phi,\rho_b(\phi))$
related by eq.~\pref{app:constraint}.

\end{document}